\documentclass[preprint,12pt]{elsarticle}

\usepackage{amssymb}

\usepackage[nodots]{numcompress}

\usepackage{graphicx}
\usepackage{graphics}
\usepackage{epsfig}
\usepackage{amsfonts}
\usepackage{amssymb}
\usepackage{amsmath}
\usepackage{amsopn}
\usepackage{hyperref}
\usepackage{lineno}
\usepackage{times}
\usepackage{xspace}
\usepackage{array}
\usepackage{booktabs}

\journal{Elsevier}

\begin{document}

\begin{frontmatter}



\title{Isogeometric nonlinear bending and buckling analysis of variable-thickness composite plate structures}

\author[1]{T. Le-Manh}
\author[2]{Q. Huynh-Van}
\author[3]{Thu D. Phan}
\author[4]{Huan D. Phan}
\author[5]{H. Nguyen-Xuan\corref{cor}}
\address[1]{Faculty of Civil Engineering, HCMC University of Technology (HUTECH), Vietnam}
\address[2]{Faculty of Mechanical Engineering, Nguyen Tat Thanh University, Vietnam}
\address[3]{Faculty of Mechanical Engineering, Ho Chi Minh City Vocational College, Vietnam}
\address[4]{Faculty of Mechanical Engineering, Ho Chi Minh City University of Technology, Vietnam}
\address[5]{Duy Tan university, Danang, Vietnam}
\cortext[cor]{Corresponding author. Email: nguyenxuanhung@duytan.edu.vn}

\begin{abstract}
This paper investigates nonlinear bending and buckling behaviours of composite plates characterized by a thickness variation. Layer interfaces are described as functions of inplane coordinates. Top and bottom surfaces of the plate are symmetric about the midplane and the plate could be considered as a flat surface in analysis along with thickness parameters which vary over the plate. The variable thickness at a certain position in the midplane is modeled by a set of control points (or thickness-parameters) through NURBS (Non-Uniform Rational B-Spline) basic functions. The knot parameter space which is referred in modelling geometry and approximating displacement variables is employed for approximating thickness, simultaneously. The use of quadratic NURBS functions results in C$^1$ continuity of modeling variable thickness and analyzing solutions. Thin to moderately thick laminates in bound of first-order shear deformation theory (FSDT) are taken into account. Strain-displacement relations in sense of von-Karman theory are employed for large deformation. Riks method is used for geometrically nonlinear analysis. The weak form is approximated numerically by the isogeometric analysis (IGA), which has been found to be a robust, stable and realistic numerical tool. Numerical results confirm the reliability and capacity of the propose method.

\end{abstract}


\end{frontmatter}


\section{Introduction}
\label{Section 1}

Composite materials are extensively employed in industry varying from large scale structures such as aircraft bodies, ship hulls to specific components of transport vehicles. Particularly, variable thickness panels are widely applicable and hence requiring appropriate approaches for analysis and design. A number of studies investigated isotropic beams and plates with the thickness change. Analytical analysis of flexural bars of uniform and variable thickness was presented by Fertis \cite{Fertis(1993)}. Ohga et al. \cite{Ohga(1995)} presented the transfer-matrix method for buckling analysis of thin-walled structures with variable thickness cross sections. Timoshenko and Woinowsky-Krieger \cite{Timoshenko(1959)} gave solutions of circular plates of nonuniform thickness in some particular cases. Wu and Liu \cite{Wu(2001)} investigated free vibration analysis of variable thickness circular plates using the generalized differential quadrature rule. Buckling analysis of rectangular plates of linearly tapered thickness was considered by Eisenberger and colleagues \cite{Eisenberger(2003),Shufri(2005)} using the extended Kantorovich method. Theories for analysis of variable thickness composite plates were also presented. Ganesan and Liu \cite{Ganesan(2008)} carried out the nonlinear buckling analysis and the prediction of the first-ply failure of tapered laminated plates under uniaxial compression. Besides, a number of works investigated laminated composite plates with thickness change caused by ply drop \cite{Dinardo(1989),Varughese(1997)}. However, most of studies considered the linear change of thickness and there is a lack of approaches for composite plates with complex nonuniform thickness. Accordingly, in this work, a numerical method considering the laminated composite plates of variable thickness is proposed based on isogeometric analysis in which the thickness of the plates is approximated from a set of control thickness parameters using NURBS basis functions.

The idea of using Non-Uniform Rational B-Spline basis functions in finite element analysis procedure was presented by Hughes et al. \cite{Hughes(2005),Cottrell(2009)} in so-called isogeometric analysis. The method filled the gap between geometric design based on NURBS (see \cite{Piegl(1997)}) and structural analysis. The smoothness and high continuity in NURBS-based finite element meshes promise more accurate solutions. Over the last decade, many efforts have been investigated in developing the method such as global refinement techniques \cite{Hughes(2005), Cottrell(2007)}, local refinement techniques \cite{Bazilevs(2010), Nguyen-Thanha(2011),Nguyen-Thanhb(2011), Kleiss(2012), Schillinger(2012)}, quadrature rules \cite{Hughes(2010), Auricchio(2012)}. Analysis of various plate structures including vibration, bending, buckling, postbuckling and nonlinear thermomechanical stability based on IGA was widely investigated in \cite{Luu(2015), Kapoor(2012), Thai(2012), Thai(2013b), Nguyen-Xuan(2013), NguyenVP(2013), Le-Manh(2014a), Le-Manh(2014b), Guo(2014a), Guo(2014b), Tran(2013a), Tran(2013b), Nguyen-Xuan(2014), Hosseini(2014), Kapoor(2013), Nguyen-Thanh(2015), Kiendl(2009), Benson(2010), Benson(2011), Echter(2013), Kiendl(2015), Deng(2015), BLieu(2016), VLTran(2016)} and so on.

Nonlinear bending and buckling analysis of variable thickness composite plates are conducted in this work. Kinematics is based on first-order shear deformation theory (FSDT) in considering the global behavior of structures. von-Karman strain-displacement relations including geometric imperfections which were presented in the works of Le-Manh and Lee \cite{Le-Manh(2014a),Le-Manh(2014b)} in investigating postbuckling of laminated composited plates are employed. Governing equations are constructed based on the virtual displacement principle of energy approach. In the isogeometric analysis framework, the 2-dimensional parameter space which is referred in constructing geometry and finite element model is also the one of thickness approximation. Surfaces of the plate could be arbitrarily defined, in assumption of being symmetric about the midplane, and then the thickness will be interpolated from a set of control thickness parameters using NURBS basis. This provides a flexible solution in modeling smooth nonuniform surfaces while plate theories are applicable. C$^1$ continuous requirements in modeling variable thickness and analyzing solutions are achieved by the use of quadratic NURBS elements. Riks algorithm is employed for geometrically nonlinear analysis. Numerical problems consisting of isotropic and laminated composite plates in various boundary conditions are carried out. Variable thickness examples are defined and examined in different amplitudes in order to demonstrate the applicability as well as accuracy of the present method.

The article is outlined beginning with a general introduction of isogeometric analysis. A brief of theoretical formulations focusing on the adjustment of kinematics and the nonuniform thickness treatment is presented in Section 3. Numerical examples describing nonlinear bending and buckling behavior of variable thickness composite plates are carried out in next Section. Some concluding remarks are drawn at last.

\section{Isogeometric analysis}
\label{Section 2}

A brief recall of NURBS-based geometry and the framework of isogeometric analysis are presented in this section.
\subsection{NURBS basis functions}
\label{Section 2.1}

A knot vector which defines the parameter space is a set of non-decreasing real numbers,

\begin{align}
\label{Eq. 1}
\Xi = \lbrace\xi_1,\xi_2,...,\xi_{n+p+1}\rbrace
\end{align}

\begin{flushleft}
where $\xi_i$ is a knot and interval $\left[ {{\xi _1},{\xi _{n + p + 1}}} \right]$ is called a patch. A knot vector is \textit{open} if the first and the last knot have the multiplicity $p+1$, and \textit{uniform} if the other knots $\xi _i$ are in uniform space.
\end{flushleft}

\begin{flushleft}
Assuming $\Xi$ is an \textit{open uniform} knot vector, B-Spline basis functions of order $p\geq 0$ are constructed using the following recursive formula \cite{Piegl(1997)},
\end{flushleft}

\begin{subequations}\label{Eq. 2}
\begin{align}
\label{Eq. 2a}
&{{N_{i,0}}(\xi ) = \left\{ {\begin{array}{*{20}{l}}
   {\begin{array}{*{20}{l}}
   1 & {{\rm{,if}}} & {{\xi _i} \le \xi  \le {\xi _{i + 1}}}  \\
\end{array}} \hfill  \\
   {\begin{array}{*{20}{l}}
   0 & {,{\rm{otherwise}}}  \\
\end{array}} \hfill  \\
\end{array}} \right.}\\
\label{Eq. 2b}
&{{N_{i,p}}(\xi ) = \frac{{\xi  - {\xi _i}}}{{{\xi _{i + p}} - {\xi _i}}}{N_{i,p - 1}}(\xi ) + \frac{{{\xi _{i + p + 1}} - \xi }}{{{\xi _{i + p + 1}} - {\xi _{i + 1}}}}{N_{i + 1,p - 1}}(\xi )}\
\end{align}
\end{subequations}

\begin{flushleft}
B-Spline basis function has important properties such as,
\end{flushleft}

\begin{itemize}
  \item Non-negative: $N_{i,p}(\xi) \geq 0, \forall\xi$,
  \item $N_{i,p}(\xi)$ is locally supported in the interval $\left[ {{\xi _i},{\xi _{i + p + 1}}} \right]$,
  \item Constitute a partition of unity: $ \sum_{i=1}^{n} N_{i,p}(\xi) = 1, \forall\xi $.
\end{itemize}

\begin{flushleft}
Considering a 2-dimensional parameter space, the basis functions are derived as,
\end{flushleft}

\begin{align}
\label{Eq. 3}
R_{ij}^{pq}\left( {\xi ,\eta } \right) = {N_{i,p}}\left( \xi  \right){M_{j,q}}\left( \eta  \right)
\end{align}

\begin{flushleft}
where $i = 1,2,...,n$ and $j = 1,2,...,m$. ${N_{i,p}}\left( \xi  \right)$ and ${M_{j,q}}\left( \eta  \right)$ are the basis of order \textit{p} and \textit{q} in $ \Xi = \lbrace \xi_1,\xi_2,...,\xi_{n+p+1} \rbrace $ and $ H  = \lbrace \eta_1,\eta_2,...,\eta_{m+q+1} \rbrace $ directions, respectively.
\end{flushleft}

\begin{flushleft}
B-Spline surface is a tensor product of a control net of n$\times$m control points $\left\{ {{B_{ij}}} \right\} \in {R^d}$ and 2-dimensional B-Spline basis functions,
\end{flushleft}

\begin{align}
\label{Eq. 4}
S\left( {\xi ,\eta } \right) = \sum\limits_{i = 1}^n {\sum\limits_{j = 1}^m {R_{ij}^{pq}\left( {\xi ,\eta } \right){B_{ij}}\left( {x,y,z} \right)} }
\end{align}

\begin{flushleft}
B-Spline surface is $ C^\infty $ continuous inside a knot span $\left[ {{\xi _i},{\xi _{i + 1}}} \right]$ and $ C^{p-1} $ continuous at a single knot $\xi_i$. NURBS surface is obtained by projecting a nonrational B-Spline surface ${S^w}\left( {\xi ,\eta } \right)$ from a homogeneous coordinate in ${R^{d + 1}}$  space onto physical ${R^d}$  space, which has a final form similar to Eq. (4) with the rational NURBS basis defined by
\end{flushleft}

\begin{align}
\label{Eq. 5}
R_{ij}^{pq}\left( {\xi ,\eta } \right) = \frac{{{N_{i,p}}\left( \xi  \right){M_{j,q}}\left( \eta  \right){w_{ij}}}}{{\sum\limits_{i = 1}^n {\sum\limits_{j = 1}^m {{N_{i,p}}\left( \xi  \right){M_{j,q}}\left( \eta  \right){w_{ij}}} } }}
\end{align}

\begin{flushleft}
where ${w_{ij}} \ge 0$  is the weight of control point $B_{ij}^w = {\left( {{w_{ij}}{x_{ij}},{w_{ij}}{y_{ij}},{w_{ij}}{z_{ij}},{w_{ij}}} \right)^T}$ in homogeneous space.
\end{flushleft}

\subsection{Isogeometric analysis model}
\label{Section 2.2}

Isogeometric analysis proposed by Hughes et al. \cite{Hughes(2005)} employs NURBS basis for geometrical modelling and finite element approximations, simultaneously. More accurate solutions compared with standard finite element analysis of composite structures are usually obtained due to the higher-order continuity in the NURBS meshes, while using less total degrees-of-freedom (DOFs). In the analysis domain, dependent displacement variables and other geometric information could be approximated by the following formula,
\begin{align}
\label{Eq. 6}
{{\bf{s}}} = \sum\limits_{c = 1}^{nC} {R_c{\bf{s}}_c}
\end{align}

\begin{flushleft}
where $nC$ is the number of control points, $ {\bf{s}}_c$ are control parameters (i.e. displacements of control points, initial deviation of control points or control thickness parameters, etc) and $R_c$ are 2-dimensional NURBS basis functions.
\end{flushleft}

\begin{flushleft}
In finite element subdomain, Eq. (6) could be written as,
\end{flushleft}

\begin{align}
\label{Eq. 7}
{{\bf{s}}^e} = \sum\limits_{c = 1}^{nCe} {R_c^e{\bf{s}}_c^e}
\end{align}

\begin{flushleft}
where $nCe$ is the number of control points per element, $ {\bf{s}}_c^e$ are control parameters of corresponding element and $R_c^e$ are NURBS basis functions evaluated with respect to parameter coordinates which define the element in physical space.
\end{flushleft}

\section{Theoretical formulation}
\label{Section 3}

In this work, first-order shear deformation is employed to consider thin and moderately thick composite plates. The kinematics of FSDT theory including geometric imperfections is carried out and constitutive relations of composite plates of variable thickness are presented in the isogeometric analysis framework.

\subsection{Kinematics}
\label{Section 3.1}

Assume that midplane is the origin of material coordinate system of a perfectly flat plate. The displacement field in FSDT is defined as follows \cite{ReddyMechLaminates(2004),HuuTai(2013)},

\begin{subequations}\label{Eq. 8}
\begin{align}
\label{Eq. 8a}
&{{u_x}(x,y,z) = u(x,y) + z{\phi _x}(x,y)}\\
\label{Eq. 8b}
&{{u_y}(x,y,z) = v(x,y) + z{\phi _y}(x,y)}\\
\label{Eq. 8c}
&{{u_z}(x,y,z) = w(x,y)}
\end{align}
\end{subequations}

\begin{flushleft}
where $\left( {{u_x},{u_y},{u_z}} \right)$  and  $\left( {u,v,w} \right)$ are the displacements of a point in body and displacements of a point on midplane along $\left( {x,y,z} \right)$  directions, respectively. ${\phi _x}$  and ${\phi _y}$  are the rotations of transverse normal of the mid-plane about the y and x axes, respectively.
\end{flushleft}

\begin{flushleft}
When the initial configuration of the plate contains geometric imperfections, i.e. initial deviations along out-of-plane direction from being perfectly flat, this small deflection can be described as a function of inplane coordinates,
\end{flushleft}
\begin{align}
\label{Eq. 9}
{{\bar{u}_z}(x,y,z) = \bar{w}(x,y)}
\end{align}

\begin{figure}[!htb]
\centering
\label{Fig. 01}
\includegraphics[width=9cm]{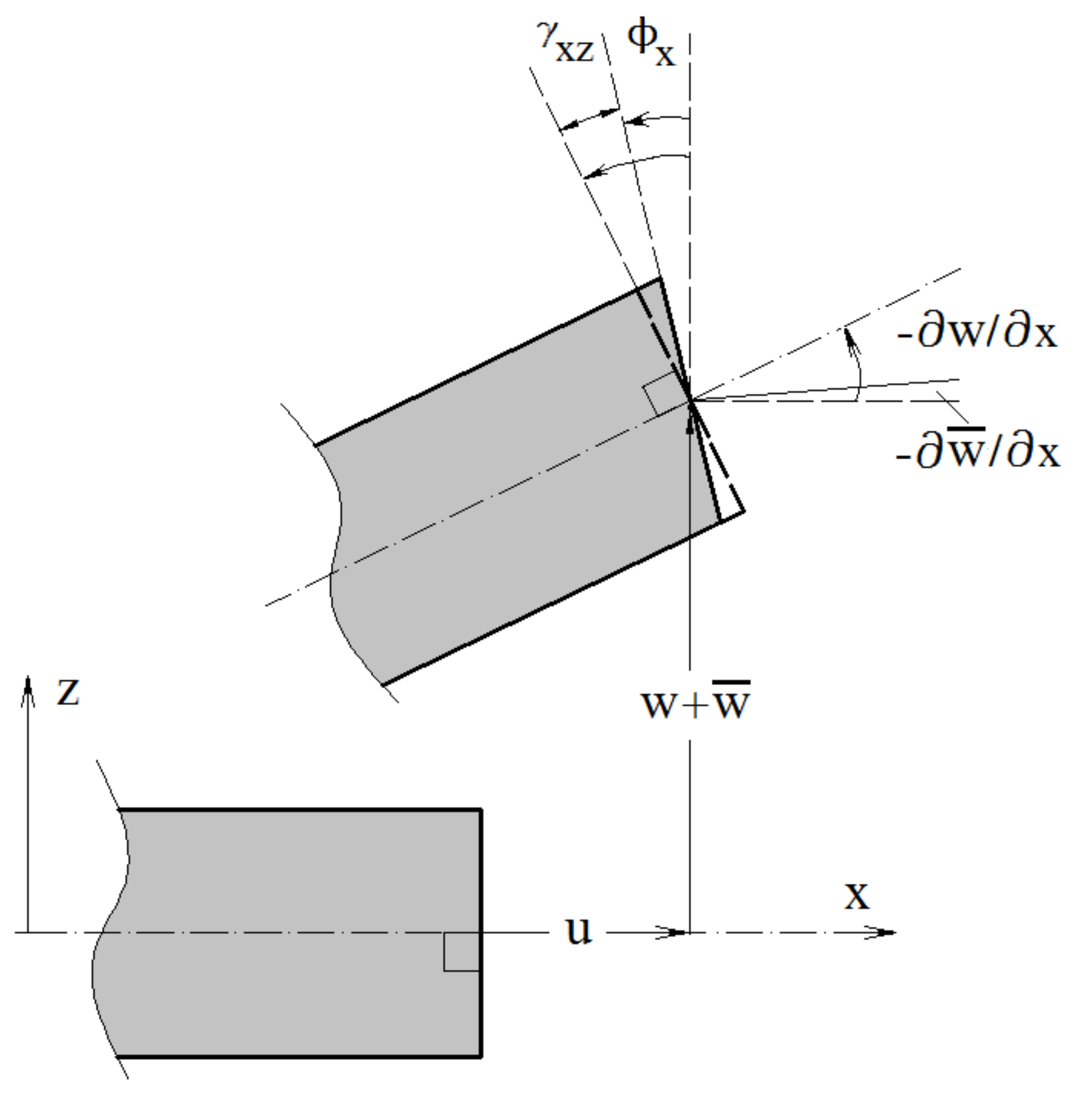}
\caption{Deformed geometry of imperfect plate in FSDT.}
\end{figure}

\begin{flushleft}
In a recent work, Le-Manh and Lee \cite{Le-Manh(2014a)} applied geometric mode shape imperfection in investigating postbuckling behavior of rectangular composite plates. The effect of imperfection magnitude was taken into account. The deformed geometry of imperfect plate in FSDT could be depicted in \hyperref[Fig. 01]{Fig. 1}. The total deflection is the summation of $w(x,y)$ and $\bar{w}(x,y)$. Nonlinear strains-displacement relations in sense of von-Karman theory including imperfections can be arranged into the inplane strains, curvatures and shear strains as follows,
\end{flushleft}
\begin{subequations}\label{Eq. 10}
\begin{align}
\label{Eq. 10a}
&{\boldsymbol{\epsilon }} = \left\{ {\begin{array}{*{20}{l}}
{\frac{{\partial u}}{{\partial x}} + \frac{1}{2}{{\left( {\frac{{\partial w}}{{\partial x}}} \right)}^2} + \frac{{\partial w}}{{\partial x}}\frac{{\partial \bar w}}{{\partial x}}}\\
{\frac{{\partial v}}{{\partial y}} + \frac{1}{2}{{\left( {\frac{{\partial w}}{{\partial y}}} \right)}^2} + \frac{{\partial w}}{{\partial y}}\frac{{\partial \bar w}}{{\partial y}}}\\
{\frac{{\partial u}}{{\partial y}} + \frac{{\partial v}}{{\partial x}} + \frac{{\partial w}}{{\partial x}}\frac{{\partial w}}{{\partial y}} + \frac{{\partial \bar w}}{{\partial x}}\frac{{\partial w}}{{\partial y}} + \frac{{\partial w}}{{\partial x}}\frac{{\partial \bar w}}{{\partial y}}}
\end{array}} \right\}\\
\label{Eq. 10b}
&{\boldsymbol{\kappa }} = \left\{ {\begin{array}{*{20}{l}}
{\frac{{\partial {\phi _x}}}{{\partial x}}}\\
{\frac{{\partial {\phi _y}}}{{\partial y}}}\\
{\frac{{\partial {\phi _x}}}{{\partial y}} + \frac{{\partial {\phi _y}}}{{\partial x}}}
\end{array}} \right\}\\
\label{Eq. 10c}
&{\boldsymbol{\gamma }} = \left\{ {\begin{array}{*{20}{l}}
{\phi {}_y + \frac{{\partial w}}{{\partial y}} + \frac{{\partial \bar w}}{{\partial y}}}\\
{{\phi _x} + \frac{{\partial w}}{{\partial x}} + \frac{{\partial \bar w}}{{\partial x}}}
\end{array}} \right\}
\end{align}
\end{subequations}

\subsection{Constitutive relations for composite plates of variable thickness}
\label{Section 3.2}

Here, composite plates whose geometry is symmetric about the midplane are taken into account, and hence their initial configuration still can be modeled as a flat surface. The thickness of the layer $k$ at a certain position on the midplane is a function of inplane coordinates $h^{(k)}(x,y)$ which could be approximated from a set of control thickness parameters $h^{(k)}_c$ using 2-dimensional NURBS basis:
\begin{align}
\label{Eq. 11}
{h^{\left( k \right)}} = \sum\limits_{c = 1}^{nC_h} {{R_c}h_c^{\left( k \right)}}
\end{align}

\begin{flushleft}
Basically, the number of control points $nC$ and the number of control thickness parameters $nC_h$ could be different, and the parameter spaces for each of them could be also separated. However, it is more convenient to use a unique parameter space in the analysis framework. Accordingly, one control thickness parameter will be assigned at each control point to define a lamina thickness. For a composite laminate made of $n$ laminae, there will be $n \times nC$ control thickness parameters in total. The thickness of the laminate is derived as follows
\end{flushleft}
\begin{align}
\label{Eq. 12}
h = \sum\limits_{k = 1}^n {{h^{\left( k \right)}}}  = \sum\limits_{k = 1}^n {\sum\limits_{c = 1}^{nC} {{R_c}h_c^{\left( k \right)}} }  = \sum\limits_{c = 1}^{nC} {{R_c}\left( {\sum\limits_{k = 1}^n {h_c^{\left( k \right)}} } \right)}
\end{align}

\begin{flushleft}
Due to the change of thickness, the extensional stiffness matrix $\textbf{A}$, the extensional-bending coupling matrix $\textbf{B}$, bending matrix $\textbf{D}$ and transverse shear stiffness matrix $\textbf{A}_s$ vary over the plate and can be computed by
\end{flushleft}
\begin{align}
\label{Eq. 13}
&{\bf{A}} = \sum\limits_{k = 1}^n {{{{\bf{\bar Q}}}^{\left( k \right)}}\left( {{{\mathord{\buildrel{\lower3pt\hbox{$\scriptscriptstyle\frown$}}
\over z} }_{k + 1}} - {{\mathord{\buildrel{\lower3pt\hbox{$\scriptscriptstyle\frown$}}
\over z} }_k}} \right)} \\
\label{Eq. 14}
&{\bf{B}} = \frac{1}{2}\sum\limits_{k = 1}^n {{{{\bf{\bar Q}}}^{\left( k \right)}}\left( {\mathord{\buildrel{\lower3pt\hbox{$\scriptscriptstyle\frown$}}
\over z} _{k + 1}^2 - \mathord{\buildrel{\lower3pt\hbox{$\scriptscriptstyle\frown$}}
\over z} _k^2} \right)} \\
\label{Eq. 15}
&{\bf{D}} = \frac{1}{3}\sum\limits_{k = 1}^n {{{{\bf{\bar Q}}}^{\left( k \right)}}\left( {\mathord{\buildrel{\lower3pt\hbox{$\scriptscriptstyle\frown$}}
\over z} _{k + 1}^3 - \mathord{\buildrel{\lower3pt\hbox{$\scriptscriptstyle\frown$}}
\over z} _k^3} \right)} \\
\label{Eq. 16}
&{{\bf{A}}_s} = {K_s}\sum\limits_{k = 1}^n {{\bf{\bar Q}}_s^{\left( k \right)}\left( {{{\mathord{\buildrel{\lower3pt\hbox{$\scriptscriptstyle\frown$}}
\over z} }_{k + 1}} - {{\mathord{\buildrel{\lower3pt\hbox{$\scriptscriptstyle\frown$}}
\over z} }_k}} \right)}
\end{align}

\begin{flushleft}
where $\bf{{\bar Q}}^{(k)}$ and $\bf{\bar Q_s} $ are the global plane-stress stiffness and shear stiffness of lamina $(k)$,  ${\mathord{\buildrel{\lower3pt\hbox{$\scriptscriptstyle\frown$}} \over z} _k} ~ (k=1,2,...,n+1)$ are the coordinates of interfaces along normal direction determined at a specific position on midplane via Gauss quadrature points in the integral loop and $K_s$ is shear correction factor.
\end{flushleft}

\begin{flushleft}
The constitutive relations are then derived  as follows,
\end{flushleft}	
\begin{align}
\label{Eq. 17}
\left\{ {\begin{array}{*{20}{c}}
{\bf{N}}\\
{\bf{M}}\\
{\bf{Q}}
\end{array}} \right\}{\bf{ = }}\left[ {\begin{array}{*{20}{c}}
{\bf{A}}&{\bf{B}}&{\bf{0}}\\
{\bf{B}}&{\bf{D}}&{\bf{0}}\\
{\bf{0}}&{\bf{0}}&{{{\bf{A}}_{{s}}}}
\end{array}} \right]\left\{ {\begin{array}{*{20}{c}}
{\boldsymbol{\epsilon }}\\
{\boldsymbol{\kappa }}\\
{\boldsymbol{\gamma }}
\end{array}} \right\}
\end{align}

\begin{flushleft}
where $\textbf{N}$, $\textbf{M}$, $\textbf{Q}$ is the vectors of inplane force resultants, moment resultants and transverse shear force resultants, respectively.
\end{flushleft}
%
%
%
%
\newpage

\section{Numerical examples}
\label{Section 4}
In this section, numerical problems including bending and buckling analysis of isotropic and composite plates of variable thickness are investigated to illustrate the application of the proposed approach. These plates vary from thin to moderately thick in bound of FSDT. The finite element meshes of full plates are constructed using quadratic NURBS elements in nonlinear parameterizations which lead to regular meshes in physical space. Reduced integration technique is employed to evaluate the transverse shear terms in Gauss quadrature.
Convergence tests are not shown in details for the sake of brevity. For illustration, 6$\times$6 quadratic NURBS elements are employed for analysis. Mesh refinement will be stated as necessary. Riks method is used for nonlinear analysis with the tolerance of displacement convergence $\delta=10^{-3}$. In buckling analysis, geometric imperfection is applied on the plates in primary mode shape with the normalized amplitude $\Delta = \bar{w}_c/ a = 10^{-5}$, (see Ref. \cite{Le-Manh(2014a)}).

\begin{flushleft}
The arc length methods for nonlinear analysis have been well known for the capacity of tracking equilibrium paths through limit point at which the tangent stiffness matrix ${{\bf{K}}_T}$ becomes singular. Among the methods, Riks algorithm is widely used. Reddy \cite{ReddyIntroNonlinear(2004)} gave an explanation of the scheme and provided a detail of computational algorithm. The residual vector is a function of both displacement $\bf{u}$ and load parameter $\lambda$: ${\bf{R}}\left( {{\bf{u}} + {\bf{\bar w}},\lambda } \right) = {\bf{K}}\left( {{\bf{u}} + {\bf{\bar w}}} \right) - \lambda {\bf{\bar F}}$; where ${\bf{\bar w}}$ is the imperfection vector, $\bf{K}$ is the direct stiffness matrix and $\bf{\bar F}$ is the preference load vector. In this study, the tolerance of displacement convergence is $\delta=10^{-3}$.
\end{flushleft}

\begin{flushleft}
The non-dimensional material properties are defined as: For isotropic plates: $E$ = 3$\times$10$^6$ and $\nu = 0.25$, and for composite plate: E$_1$/E$_2$ = 25, G$_{12}$/E$_2$ = 0.5, G$_{23}$/E$_2$ = 0.2, G$_{13}$ = G$_{12}$ and $\nu = 0.25$.
\end{flushleft}
\subsection{Isotropic square plates of linearly tapered thickness along x direction}
\label{Example 4.1}
\begin{figure}[!htb]
\centering
\label{Fig. 02}
\includegraphics[width=9cm]{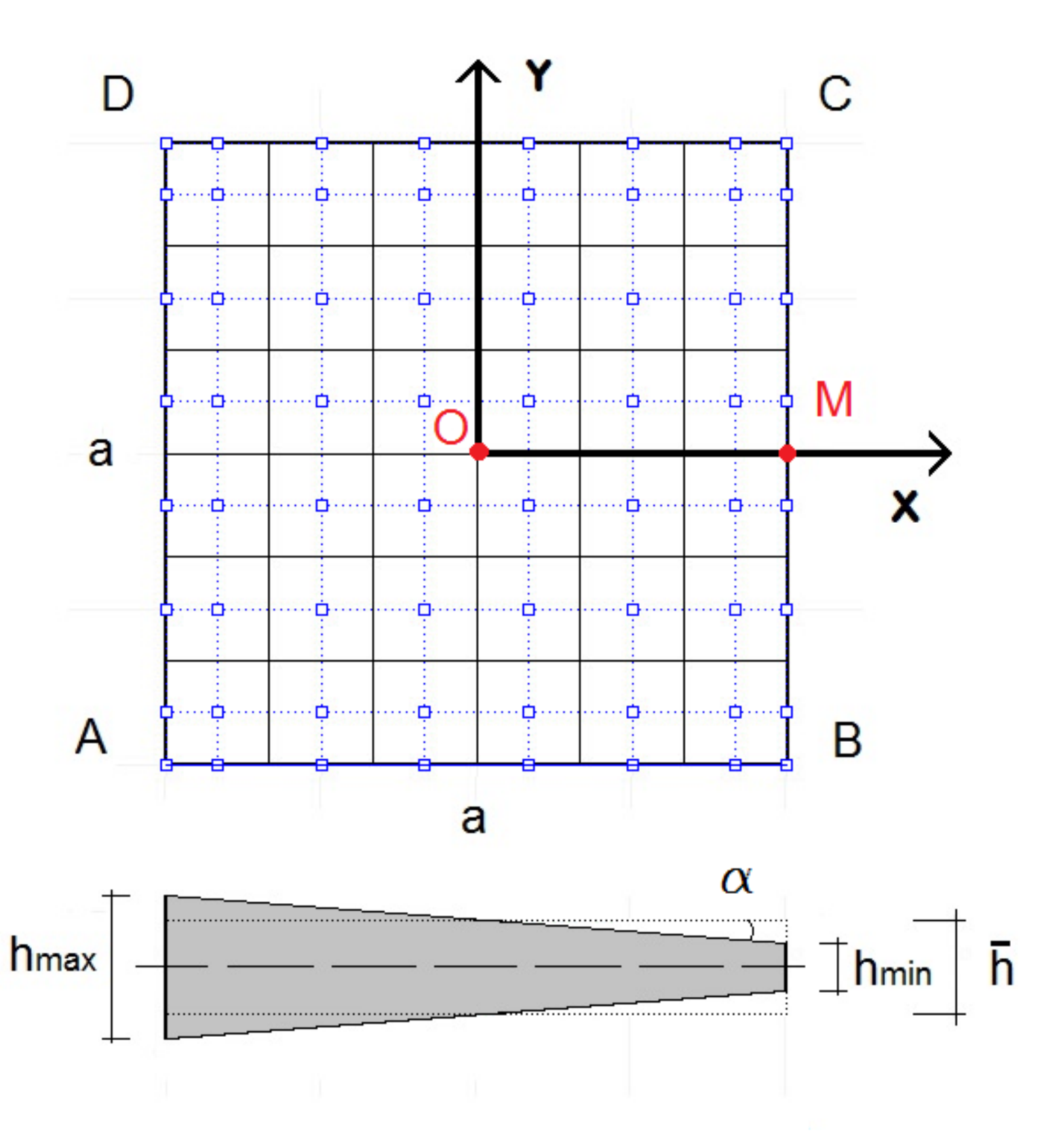}
\caption{Model of a square plate of linearly tapered thickness along $x$ direction.}
\end{figure}

The model of a square plate with the thickness h linearly decreasing along x direction is depicted in \hyperref[Fig. 02]{Fig. 2}. Inplane dimension of the plate is $a\times a$. The tapered ratio $\alpha$, i.e. the slope of the tapered surface, is defined by
\begin{align}
\label{Eq. 18}
 \alpha  = \frac{{{h_{\max }} - \bar h}}{a} = \frac{{\bar h - {h_{\min }}}}{a}
 \end{align}

\begin{flushleft}
where $\bar h$ is the uniform thickness, $\alpha=0$, of the plate which has the same volume with those of tapered thickness. The top surface of the plate could be described as the following function
\end{flushleft}
 \begin{align}
 \label{Eq. 19}
 z(x) = \frac{{\bar h}}{2} - \alpha x
 \end{align}

\begin{flushleft}
Moreover, the bottom surface is geometrically symmetric with the top surface about the midplane. In this example, isotropic square plates are investigated. The non-dimensional geometry and uniform thickness are assumed as $a$=10 and $\bar h$=0.2 (a/$\bar h$ = 50). In bending test, the plate is clamped at AD and subjected to increasing uniform transverse load, deflection of point M is measured. In buckling test, all 4 edges of the plate is simply-supported (SS1), uniaxial compression load is applied in $x$ direction and central transverse displacement at origin $O$ is measured:
\end{flushleft}
\begin{align}
\label{Eq. 20}
{\rm{Clamped~at~AD:~~}}u = v = w = {\phi _x} = {\phi _y} = 0
\end{align}

\begin{subequations}\label{Eq. 21}
\begin{align}
\label{Eq. 21a}
&{\rm{SS1~at~AD~and~BC:~~}}w = {\phi _y} = 0\\
\label{Eq. 21b}
&{\rm{SS1~at~AB~and~CD:~~}}w = {\phi _x} = 0
\end{align}
\end{subequations}

\begin{flushleft}
The transverse displacement $w$, uniform transverse load $q$ and compression load $N$ are normalized as
\end{flushleft}
\begin{align}
\label{Eq. 22}
&\bar w = \frac{w}{{\bar h}}\\
\label{Eq. 23}
&\bar q = \frac{{q{{a}^4}}}{{E{{\bar{h}}^4}}}\\
\label{Eq. 24}
&\lambda  = \frac{{N{a^2}}}{{{\pi ^2}{\bar{D}}}}
\end{align}

\begin{flushleft}
where $\bar{D}$ is  the flexural rigidity of the plate of uniform thickness , i.e. tapered ratio $\alpha=0$.
\end{flushleft}

\begin{figure}[!htb]
\centering
\label{Fig. 03}
\includegraphics[width=9cm]{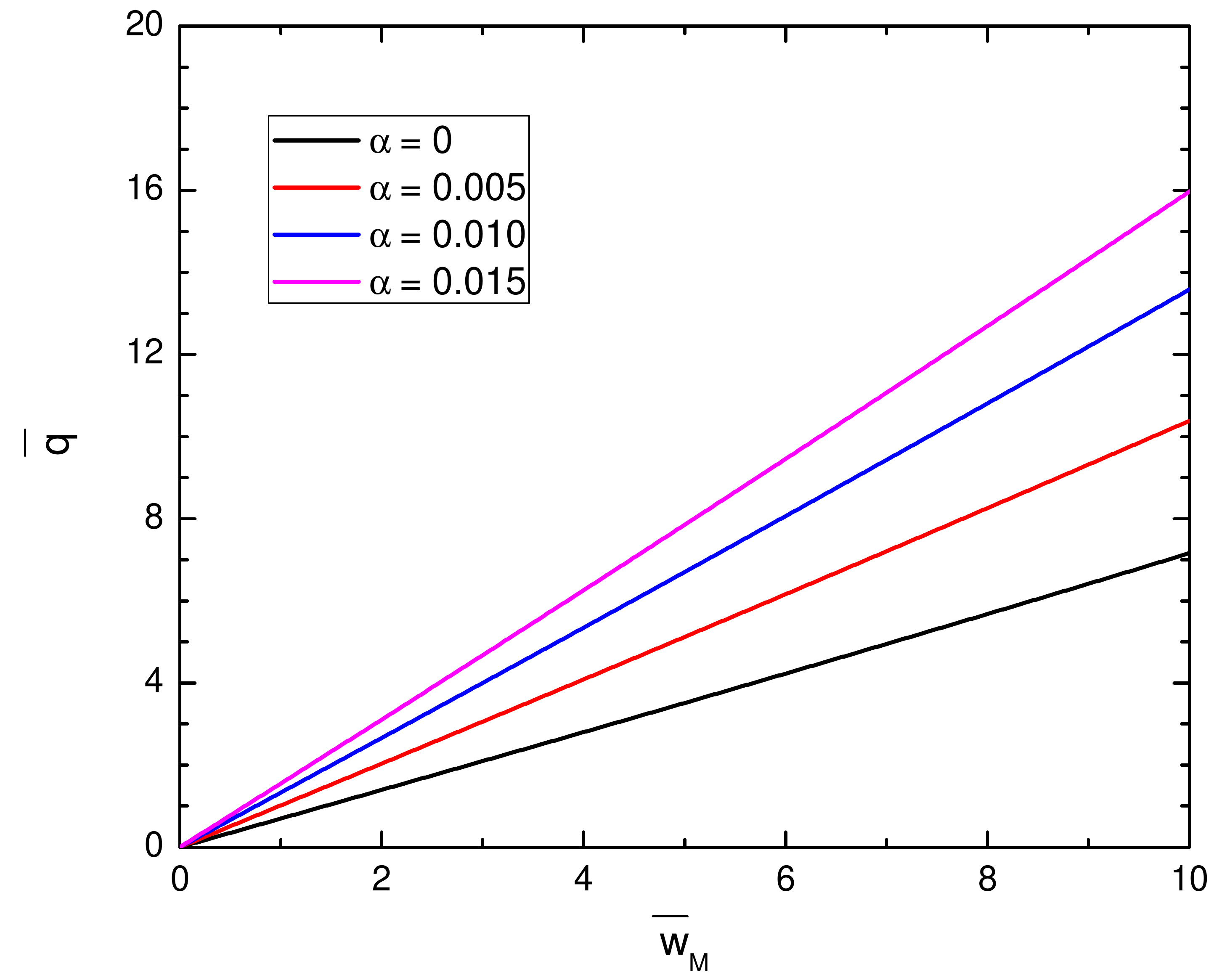}
\caption{Deflection versus uniform transverse load of a cantilever isotropic square plate of tapered thickness along $x$ direction with different tapered ratio $\alpha$.}
\end{figure}

\begin{figure}[!htb]
\centering
\label{Fig. 04}
\includegraphics[width=9cm]{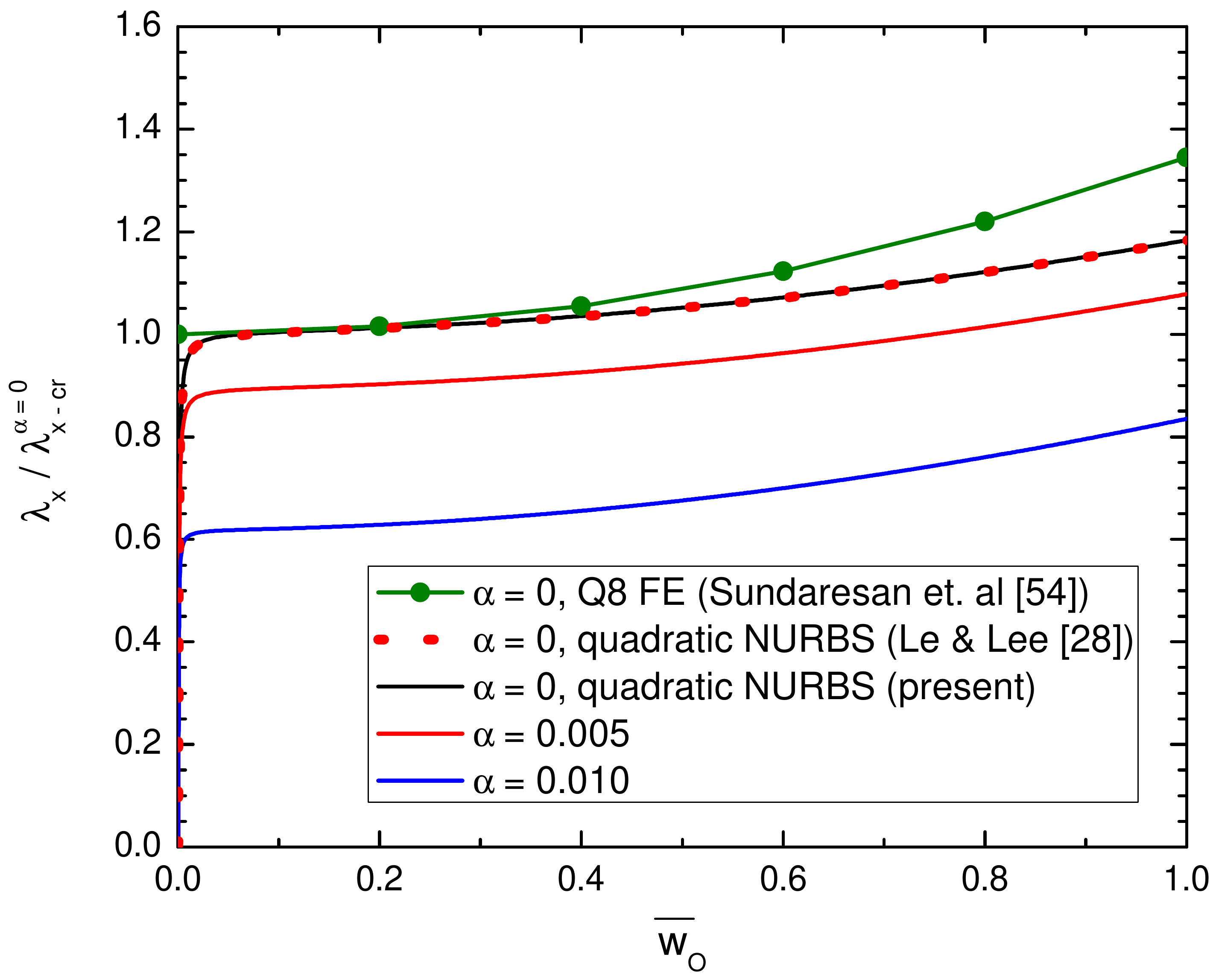}
\caption{Buckling and postbuckling path of a simply-supported isotropic square plate of tapered thickness along $x$ direction with different tapered ratio $\alpha$ under uniaxial compression load in $x$ direction.}
\end{figure}

\begin{flushleft}
The solutions are given in \hyperref[Fig. 03]{Fig. 3} and \hyperref[Fig. 04]{Fig. 4}. In bending test, linear responses of the cantilever plates under uniform transverse load are obtained. It is also reasonable that the larger tapered thickness ratio $\alpha$ results in smaller deflection. \hyperref[Fig. 04]{Fig. 4} presents buckling and postbuckling path of imperfect isotropic square plates under uniaxial compression load in x direction. The results are in good agreement with ones of uniform thickness $\alpha=0$ given by Sundaresan et al. \cite{Sundaresan(1996)} and Le-Manh and Lee \cite{Le-Manh(2014a)}. It can be seen that when increasing the slope of tapered surfaces, the critical load of the plate significantly decreases. This indicates that stability analysis should be taken into account in designing tapered structures.
\end{flushleft}

\subsection{Isotropic square plates of linearly tapered thickness along the diagonal direction}
\label{Example 4.2}

\begin{figure}[!htb]
\centering
\label{Fig. 05}
\includegraphics[width=11cm]{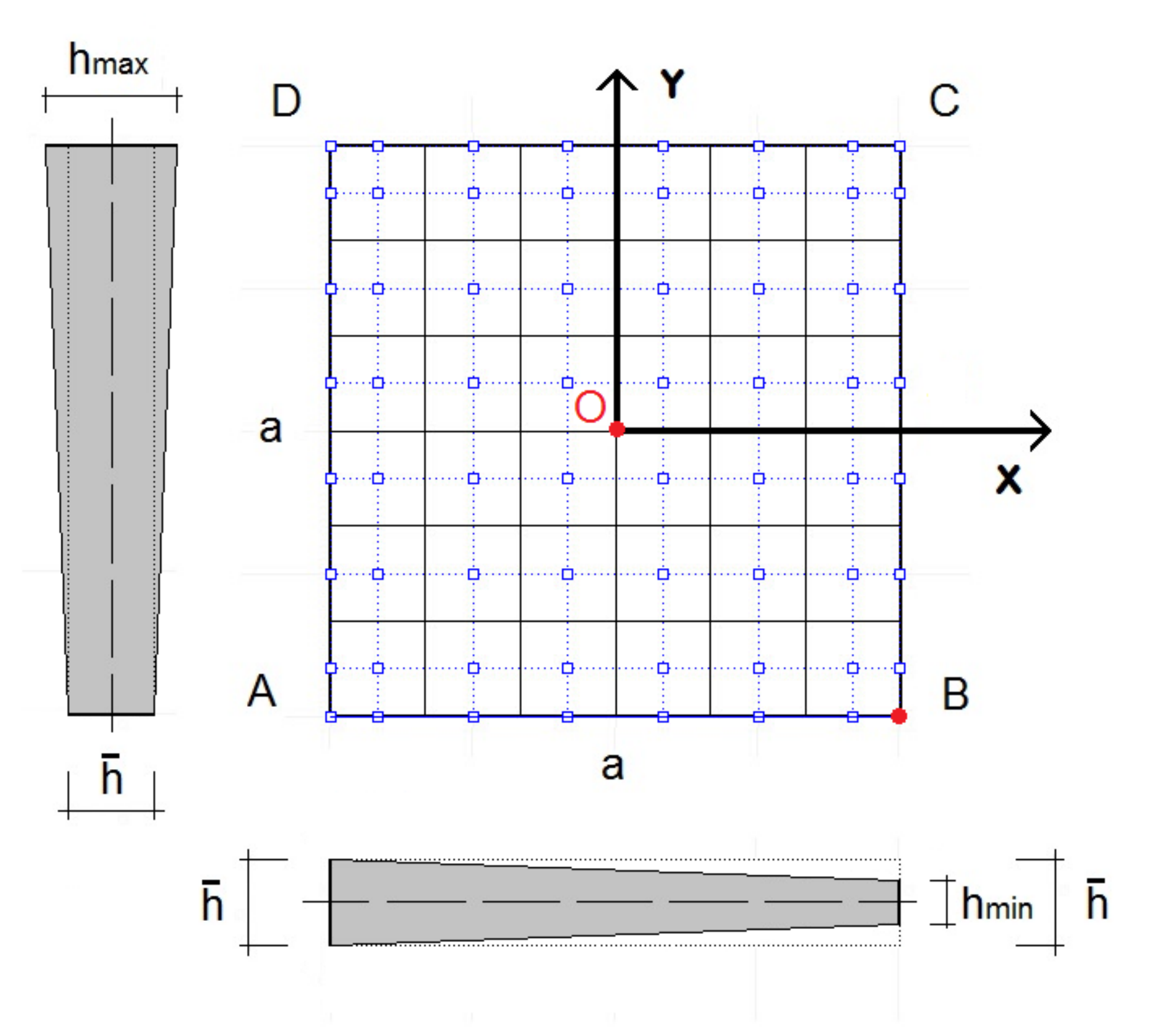}
\caption{Model of a square plate of linearly tapered thickness along diagonal direction.}
\end{figure}

In this problem, isotropic square plates of tapered thickness in two dimensions are investigated. The thickness h is gradually changes along the diagonal DB of the plate, \hyperref[Fig. 05]{Fig. 5}. In this case, the tapered ratio $\alpha$ and the top surface of the plate could be derived as follows
\begin{align}
\label{Eq. 25}
&\alpha  = \frac{{{h_{\max }} - \bar h}}{{a\sqrt 2 }} = \frac{{\bar h - {h_{\min }}}}{{a\sqrt 2 }}\\
\label{Eq. 26}
& z(x,y) = \frac{{\bar h}}{2} - \frac{\alpha }{{\sqrt 2 }}x + \frac{\alpha }{{\sqrt 2 }}y
\end{align}

\begin{flushleft}
It should be noted the plates with different tapered ratios have the same volume. Non-dimensional geometry and normalized parameters are similar to the previous example. In bending test, edge AD and CD are under simply-supported (SS2) in Eq. \eqref{Eq. 27}. The plate is subjected to increasing uniform transverse load and corresponding transverse displacement is computed at the corner point $B$. In buckling test, boundary conditions SS1 in Eq. \eqref{Eq. 21} are applied and the plate is under biaxial compression loads in $x$ and $y$ directions:
\end{flushleft}
\begin{align}
\label{Eq. 27}
{\rm{SS2~at~AD~and~DC:~~}}u = v = w = 0
\end{align}

\begin{figure}[!htb]
\centering
\label{Fig. 06}
\includegraphics[width=9cm]{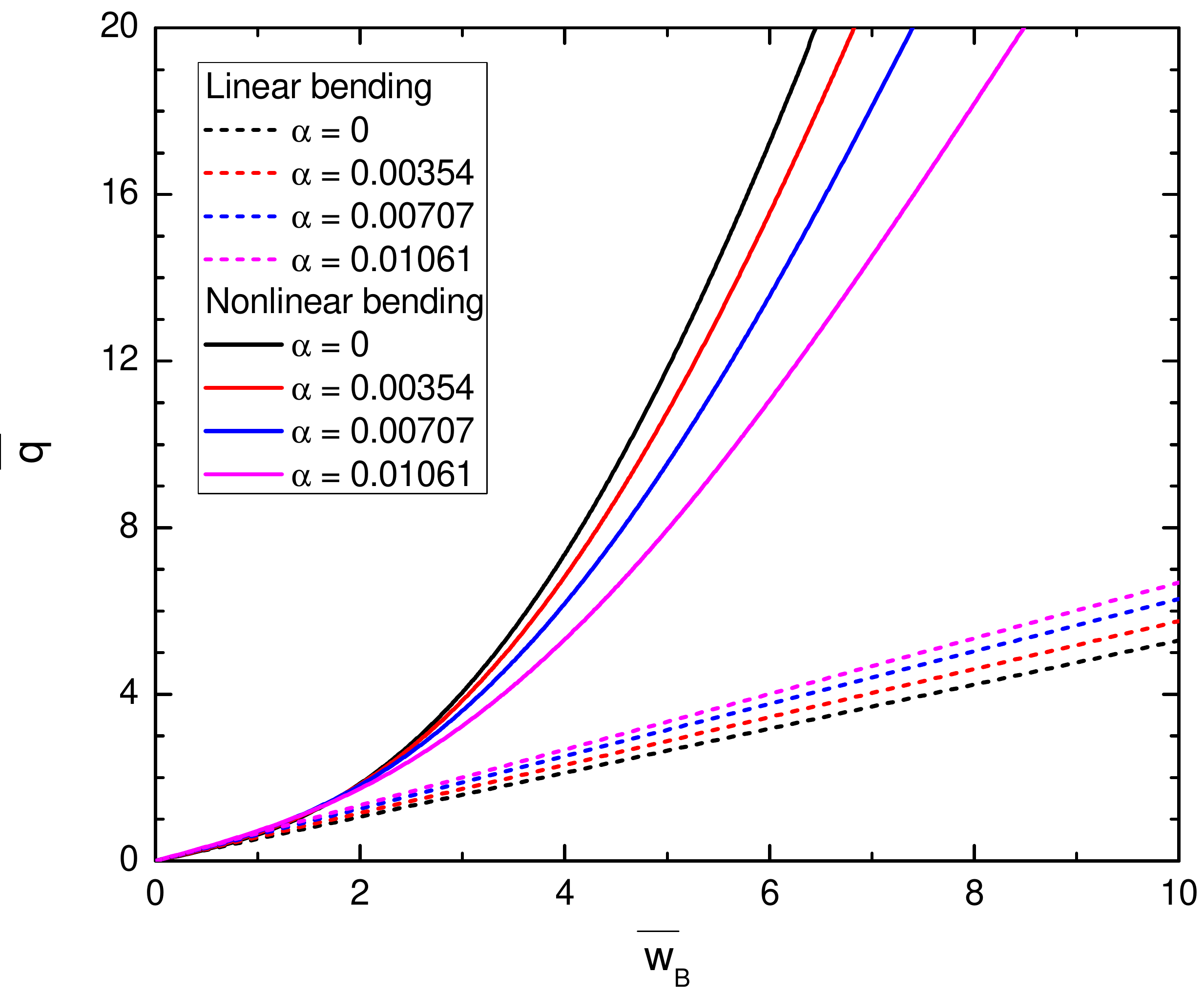}
\caption{Deflection versus uniform transverse load of an isotropic square plate of tapered thickness along diagonal direction with different tapered ratio $\alpha$.}
\end{figure}

\begin{figure}[!htb]
\centering
\label{Fig. 07}
\includegraphics[width=9cm]{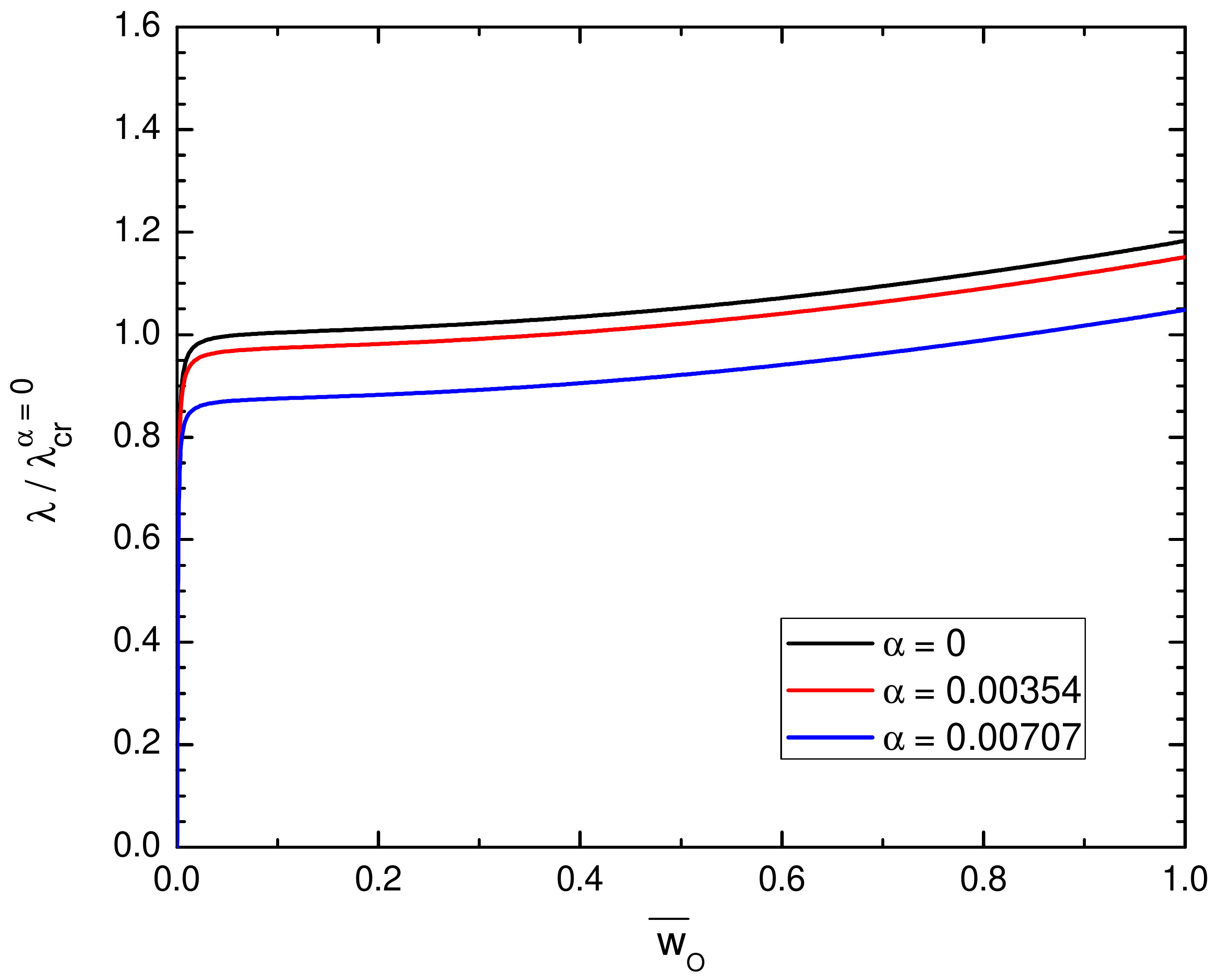}
\caption{Buckling and postbuckling path of an isotropic square plate of tapered thickness along diagonal direction with different tapered ratio $\alpha$ under biaxial compression loads in $x$ and $y$ direction.}
\end{figure}

\begin{flushleft}
The results of bending test are shown in \hyperref[Fig. 06]{Fig. 6}. It is interesting to observe the different behavior in linear and nonlinear analysis. When increasing the tapered ratio $\alpha$ the linear bending deflection decreases, however, when the transverse displacement is considerably large, higher tapered ratio results in larger nonlinear bending deflection. Buckling and postbuckling path of the isotropic square plates under biaxial compression loads are plotted in \hyperref[Fig. 07]{Fig. 7}. The critical buckling load notably decreases when the slope of tapered surfaces increases.
\end{flushleft}

\subsection{Symmetric cross-ply (0/90)$_s$ laminated composite square plate of linearly tapered thickness along x direction}

\begin{figure}[!htb]
\centering
\label{Fig. 08}
\includegraphics[width=10cm]{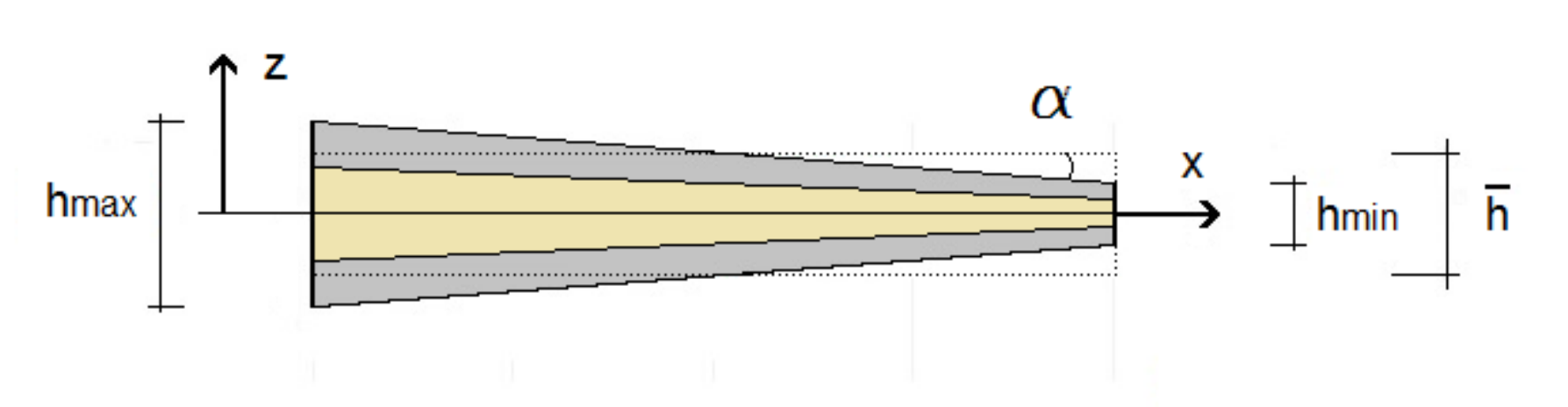}
\caption{Cross section of a symmetric cross-ply (0/90)$_s$ laminated composite plate of linearly tapered thickness along $x$ direction in $x-z$ plane.}
\end{figure}

Considering a cantilever symmetric cross-ply (0/90)$_s$ square laminated composite plate of tapered thickness along $x$ direction. The non-dimensional inplane dimension and uniform thickness of the plate are $a$=10 and $\bar h$=0.2 (a/$\bar h$ = 50). Cross section of the laminate in $x-z$ plane is depicted in \hyperref[Fig. 08]{Fig. 8}. The thickness of each ply is equal together at any position over the plate. Boundary and load conditions for bending and buckling analysis are similar to example 4.1. Uniform transverse load $q$ and compression load $N$ are normalized as follows,
\begin{align}
\label{Eq. 28}
&\bar q = \frac{{q{{a}^4}}}{{{E_2}{{\bar{h}}^4}}}\\
\label{Eq. 29}
&\lambda  = \frac{{N{a^2}}}{{{E_2}{\bar{h}^3}}}
\end{align}

\begin{figure}[!htb]
\centering
\label{Fig. 09}
\includegraphics[width=9cm]{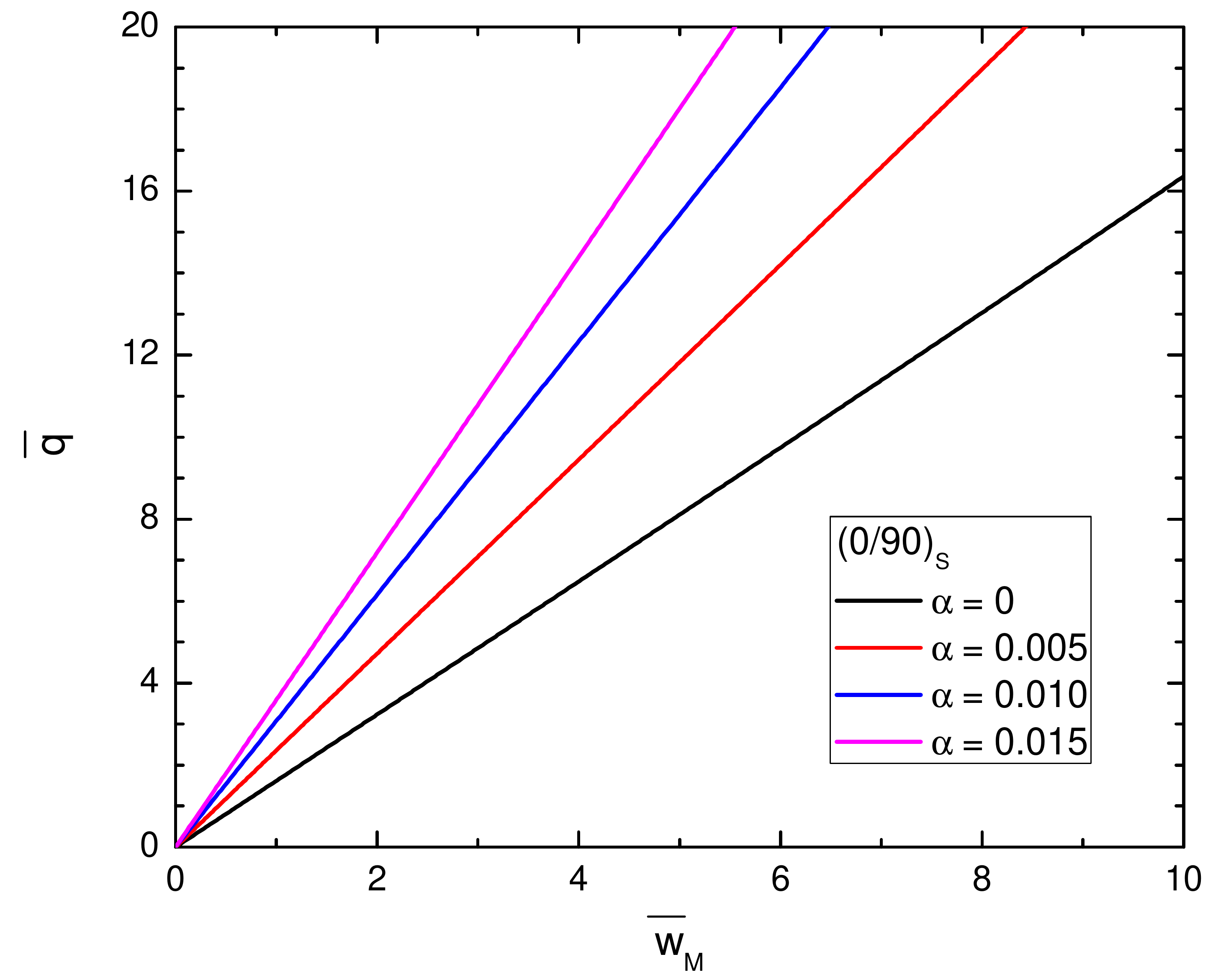}
\caption{Deflection versus uniform transverse load of a symmetric cross-ply (0/90)$_s$ square laminate of tapered thickness along $x$ direction with different tapered ratio $\alpha$.}
\end{figure}

\begin{figure}[!htb]
\centering
\label{Fig. 10}
\includegraphics[width=9cm]{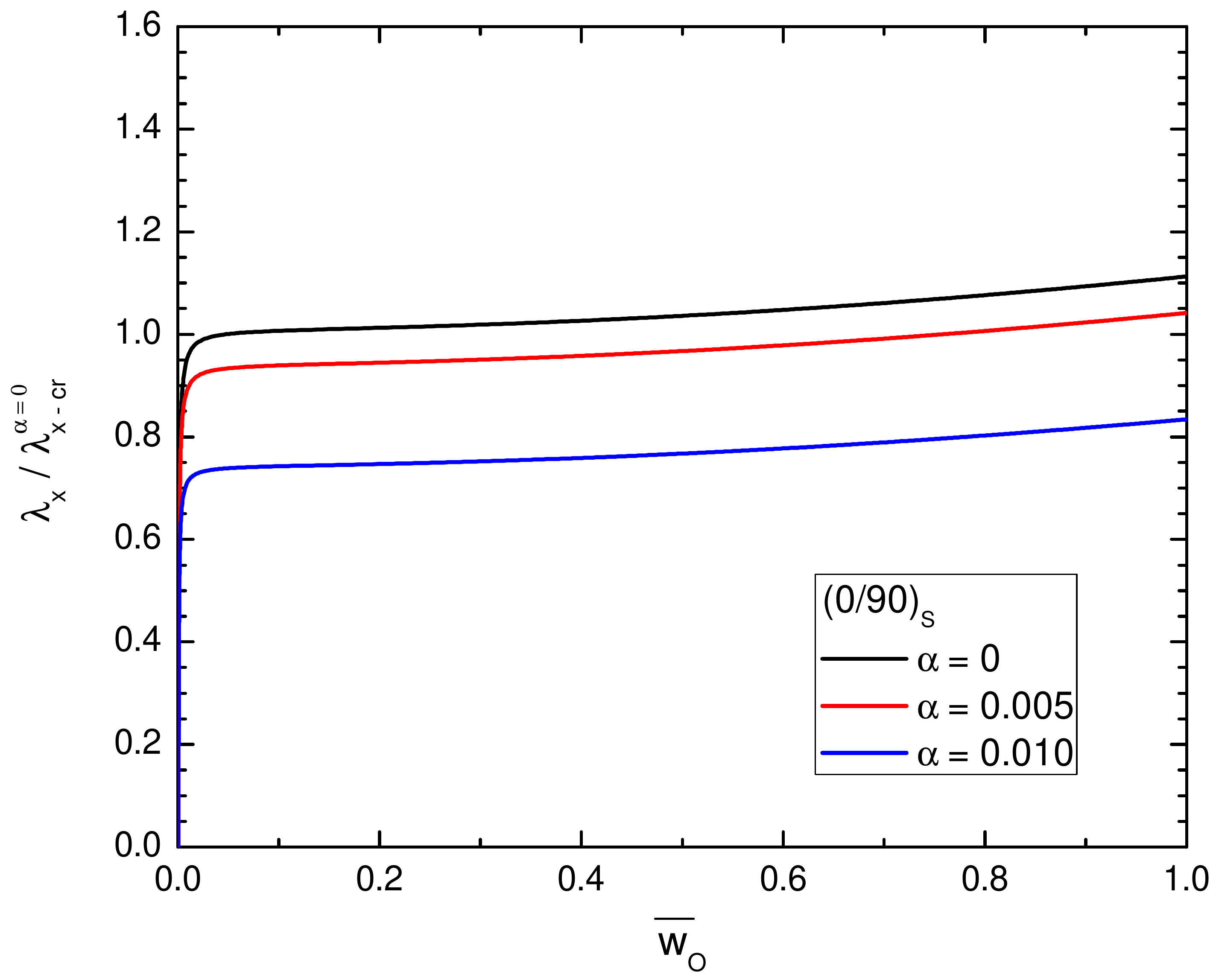}
\caption{Buckling and postbuckling path of symmetric cross-ply (0/90)$_s$ square laminate of tapered thickness along $x$ direction with different tapered ratio $\alpha$ under uniaxial compression load in $x$ direction.}
\end{figure}

\begin{flushleft}
The obtained transverse displacement versus loading curves are shown in \hyperref[Fig. 09]{Fig. 9} for bending analysis and \hyperref[Fig. 10]{Fig. 10} for buckling analysis. Behavior of the cantilever symmetric cross-ply laminate has the similar pattern with the clamped isotropic plate. The deflection of the cantilever composite plate decreases when the tapered ratio $\alpha$ increases, and the buckling load also drops considerably.
\end{flushleft}

\subsection{Symmetric angle-ply (45/-45)$_s$ laminated composite square plate of linearly tapered thickness along the diagonal direction}

A symmetric angle-ply (45/-45)$_s$ square plate of linearly tapered thickness along diagonal direction (see \hyperref[Fig. 05]{Fig. 5}) is taken in account here. Boundary and load conditions are applied same in example 4.2. The thickness of each ply is equal together at any position over the plate. The inplane dimensions, uniform thickness and normalized parameters are similar to example 4.3.

\begin{figure}[!htb]
\centering
\label{Fig. 11}
\includegraphics[width=9cm]{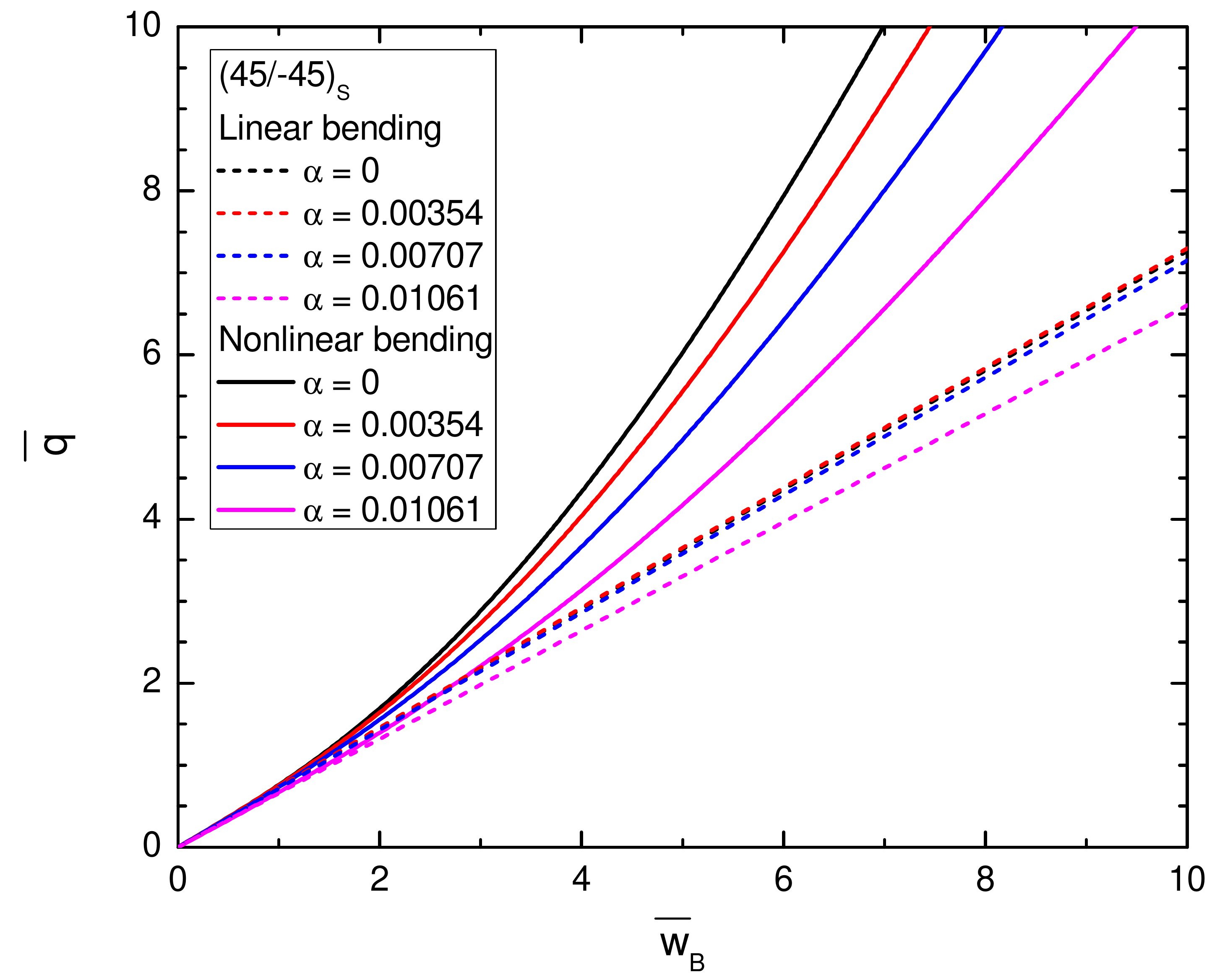}
\caption{Deflection versus uniform transverse load of a symmetric angle-ply (45/-45)$_s$ square laminate of tapered thickness along the diagonal direction with different tapered ratio $\alpha$.}
\end{figure}

\begin{figure}[!htb]
\centering
\label{Fig. 12}
\includegraphics[width=9cm]{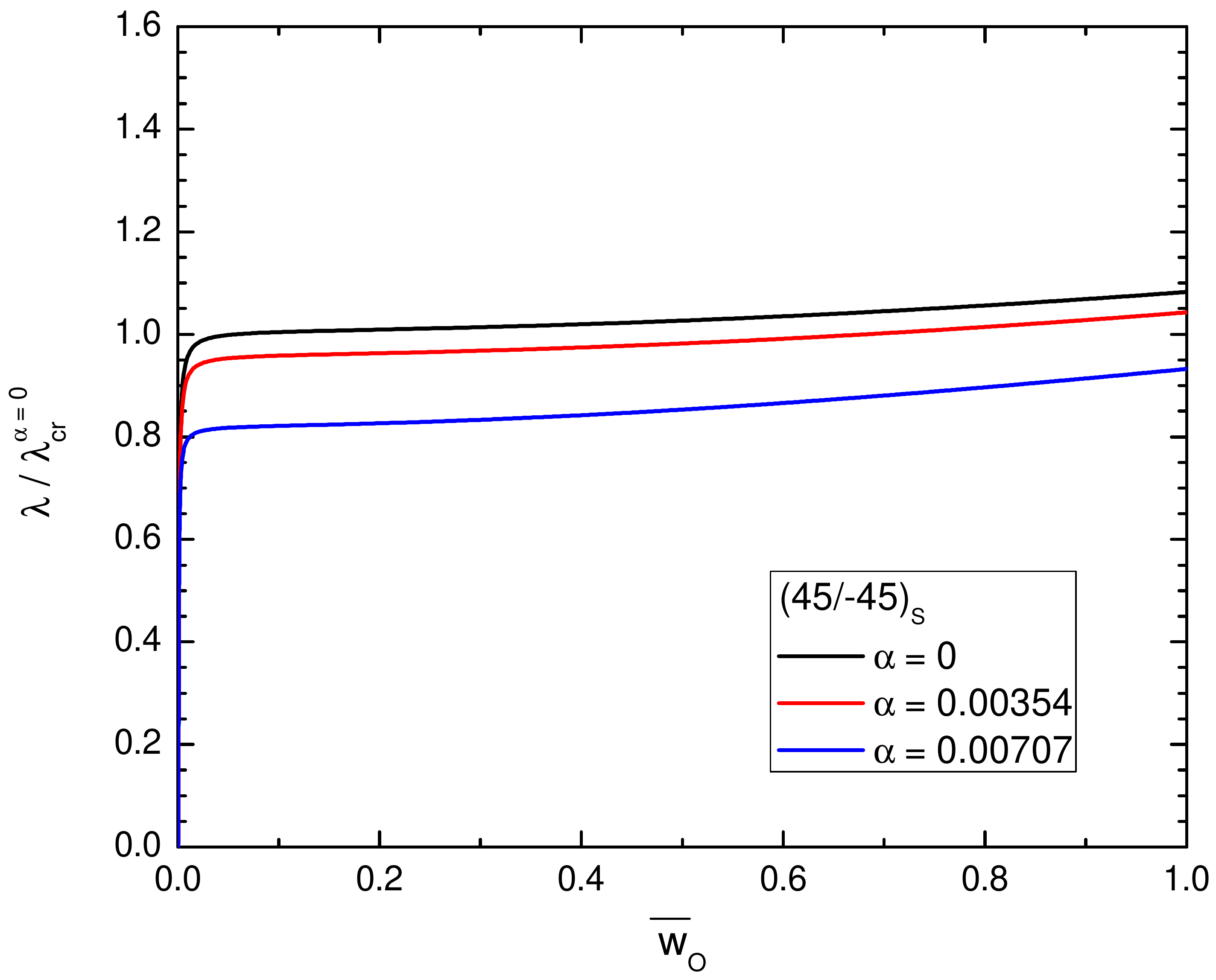}
\caption{Buckling and postbuckling path of a symmetric angle-ply (45/-45)$_s$ square laminate of tapered thickness along the diagonal direction with different tapered ratio $\alpha$ under biaxial compression load in $x$ and $y$ directions.}
\end{figure}

\begin{flushleft}
The results of bending test are given in \hyperref[Fig. 11]{Fig. 11}. It can be seen that in linear bending analysis there is an unpredictable development of the deflection with respect to the increase of tapered ratio $\alpha$. Therefore, nonlinear analysis should be considered in designing plates of tapered thickness, especially in cases of complex boundary conditions. Besides, critical load of the symmetric angle-ply (45/-45)$_s$ laminate under biaxial compression loads notably drops when increasing the slope of the tapered surface, see \hyperref[Fig. 12]{Fig. 12}.
\end{flushleft}

\subsection{Composite square plates of thickness as a sine wave}
\begin{figure}[!htb]
\centering
\label{Fig. 13}
\includegraphics[width=9cm]{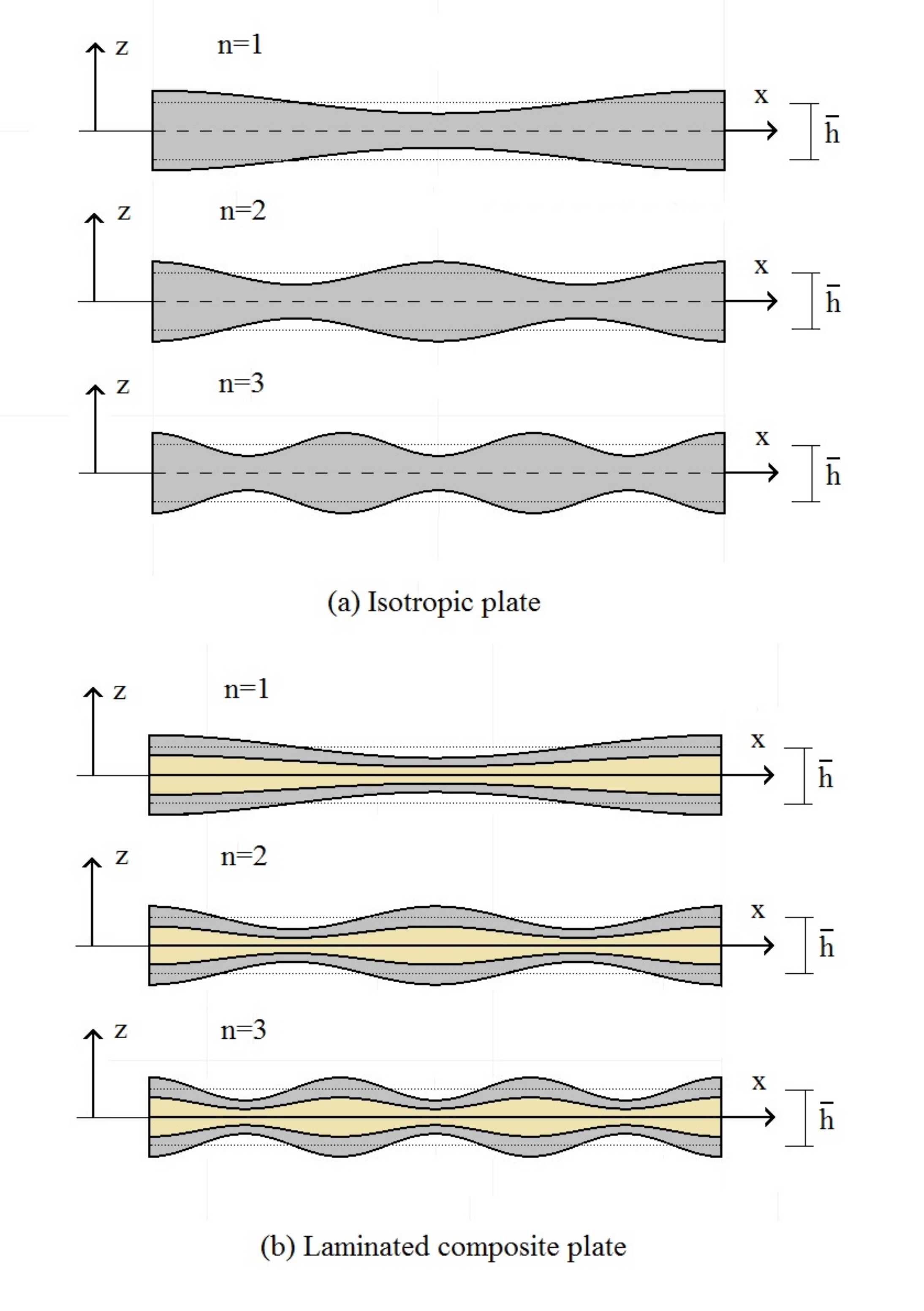}
\caption{Cross section of the plates in $x-z$ plane}
\end{figure}

In this example, nonlinear buckling of composite plates including isotropic square plate and symmetric cross-ply (0/90)$_s$ laminated composite square plates of variable thickness is examined. The top surface of the plates is described as a sine wave function, Eq. \eqref{Eq. 30}, and the bottom surface is geometrically symmetric with the top surface about the midplane. The cross section of the plates are illustrated in \hyperref[Fig. 13]{Fig. 13}. Non-dimensional inplane dimension and the uniform thickness are a=10 and $\bar h$=0.5 (a/$\bar h$=20).
\begin{align}
\label{Eq. 30}
z(x) = \alpha {\bar {h} }\sin {\rm{(}}2\pi \frac{{nx}}{a}{\rm{ + }}\frac{\pi }{2}{\rm{)  +  }}\frac{{{{\bar h}}}}{2}
\end{align}

\begin{flushleft}
where $\alpha$ is the parameter that controls the amplitude of the sine wave and $n$ is to change the number of wavelength. The plates are subjected to uniaxial compression load in y direction. When increasing the number of wavelength, the plate surfaces are more complex and it is required mesh refinement to obtain convergence solution. In this works, the mesh of 6$\times$6 quadratic NURBS elements is used for $\alpha=0$, 12$\times$12 quadratic NURBS elements for $\alpha\neq0$ and $n$=1,2 and 3.
\end{flushleft}

\begin{figure}[!htb]
\centering
\label{Fig. 14}
\includegraphics[width=9cm]{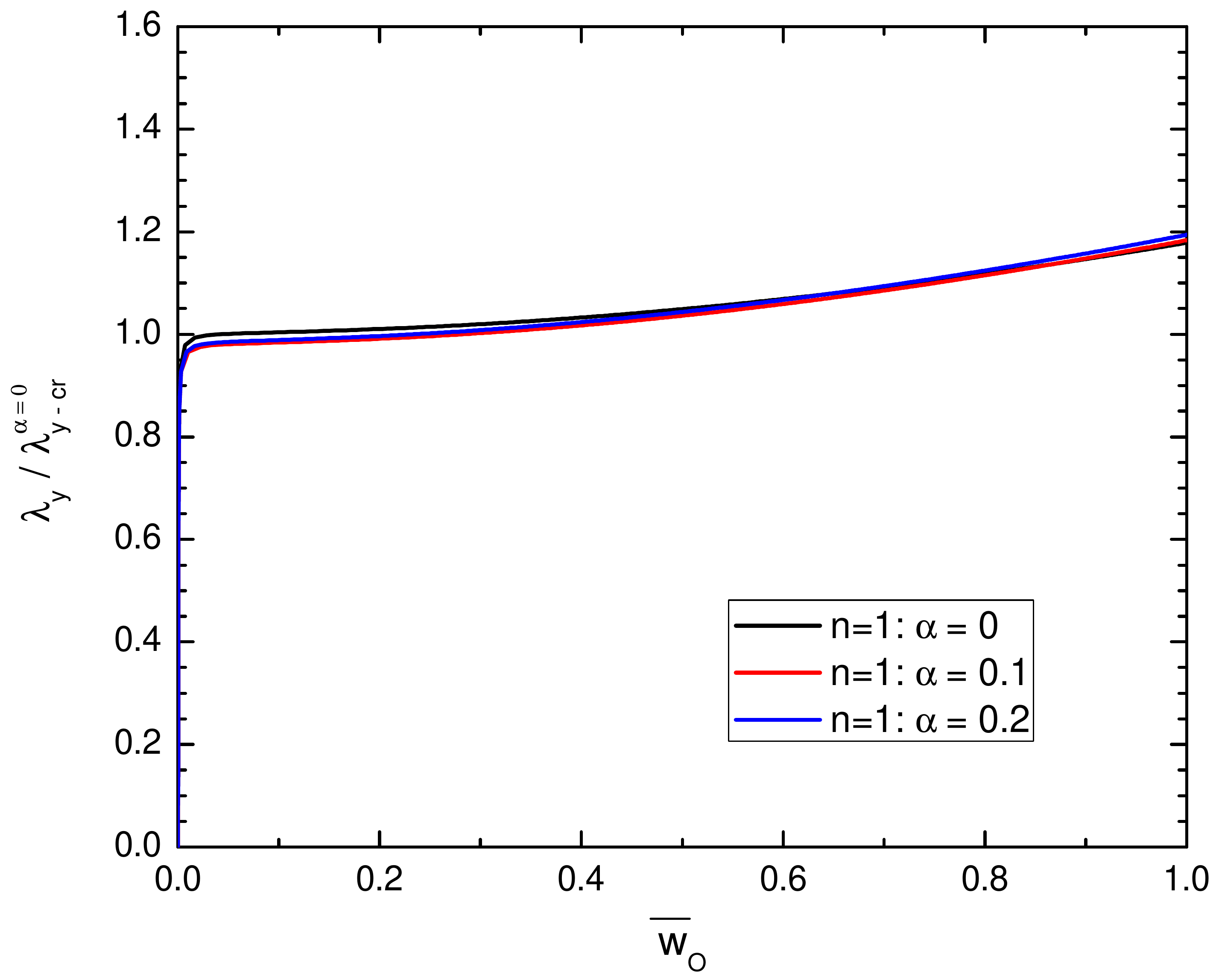}
\caption{Buckling and postbuckling path of the isotropic square plate of sine-wave thickness with fixed the wavelength $n=1$ and varying amplitude $\lambda$ under uniaxial compression load in $y$ direction.}
\end{figure}

\begin{figure}[!htb]
\centering
\label{Fig. 15}
\includegraphics[width=9cm]{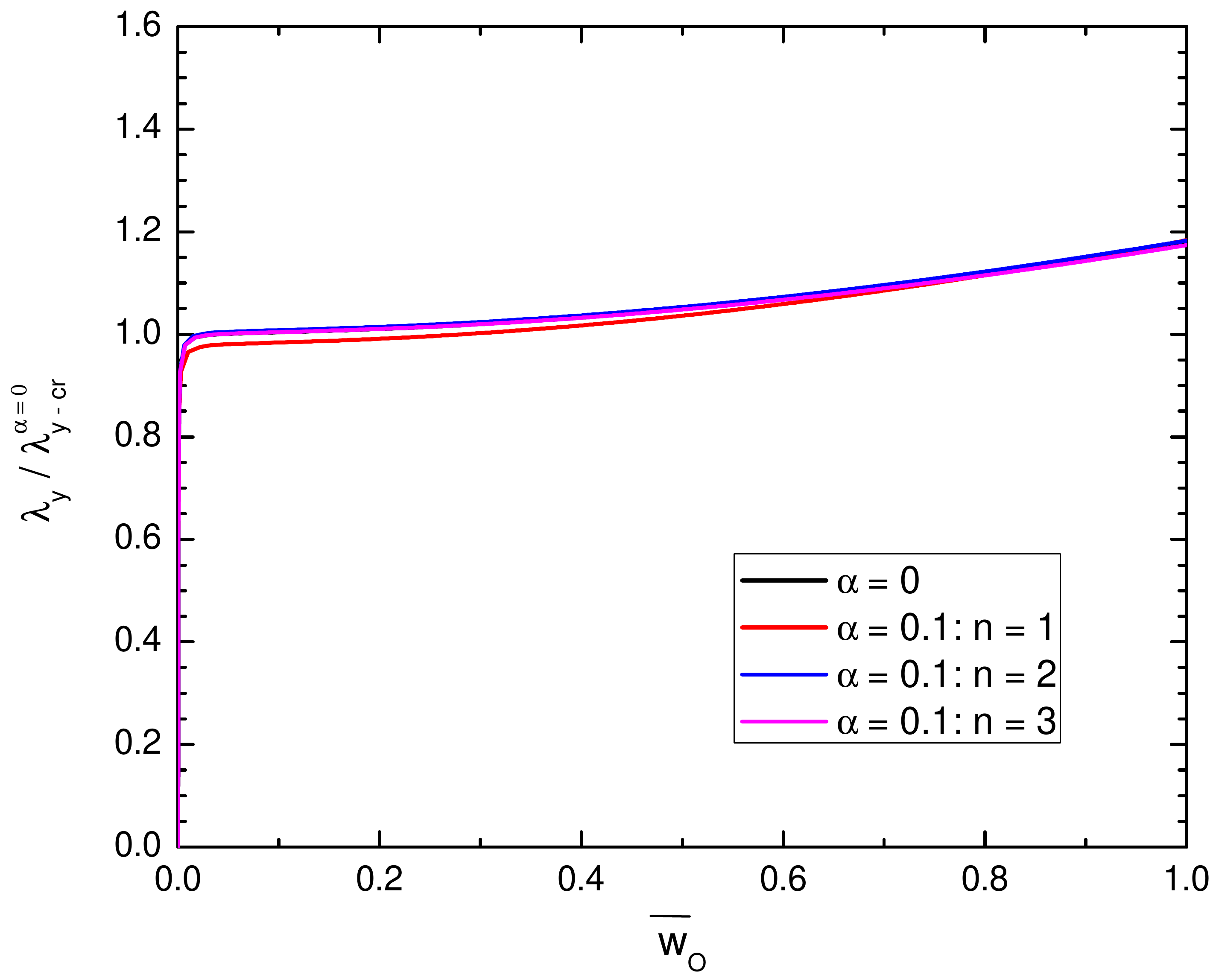}
\caption{Buckling and postbuckling path of the isotropic square plate of sine-wave thickness with fixed amplitude $\lambda$=0.1 and varying the wavelength $n$ under uniaxial compression load in $y$ direction.}
\end{figure}

\begin{figure}[!htb]
\centering
\label{Fig. 16}
\includegraphics[width=9cm]{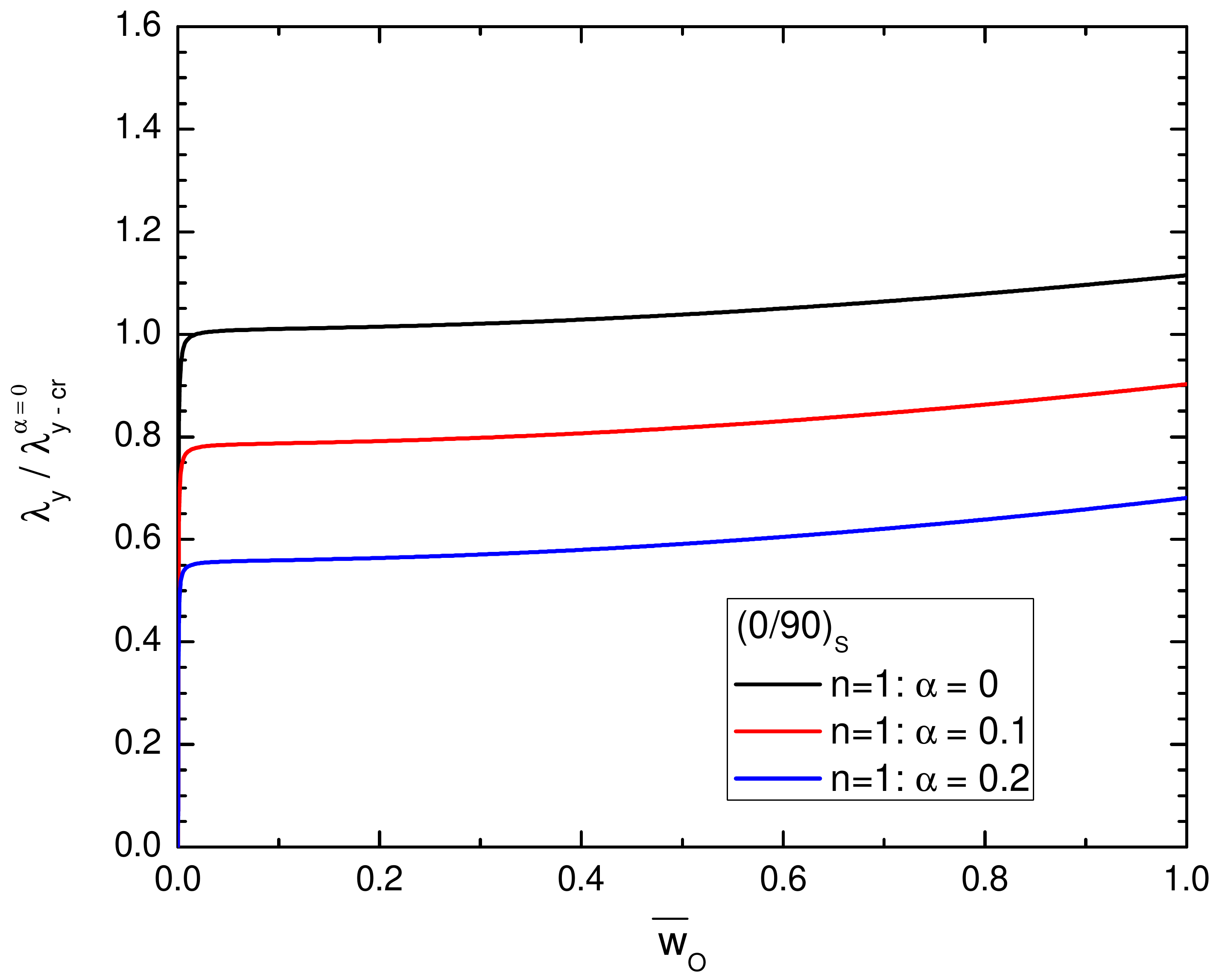}
\caption{Buckling and postbuckling path of symmetric cross-ply (0/90)$_s$ laminated composite square plate of sine-wave thickness with fixed the wavelength $n=1$ and varying amplitude $\lambda$ under uniaxial compression load in $y$ direction.}
\end{figure}

\begin{figure}[!htb]
\centering
\label{Fig. 17}
\includegraphics[width=9cm]{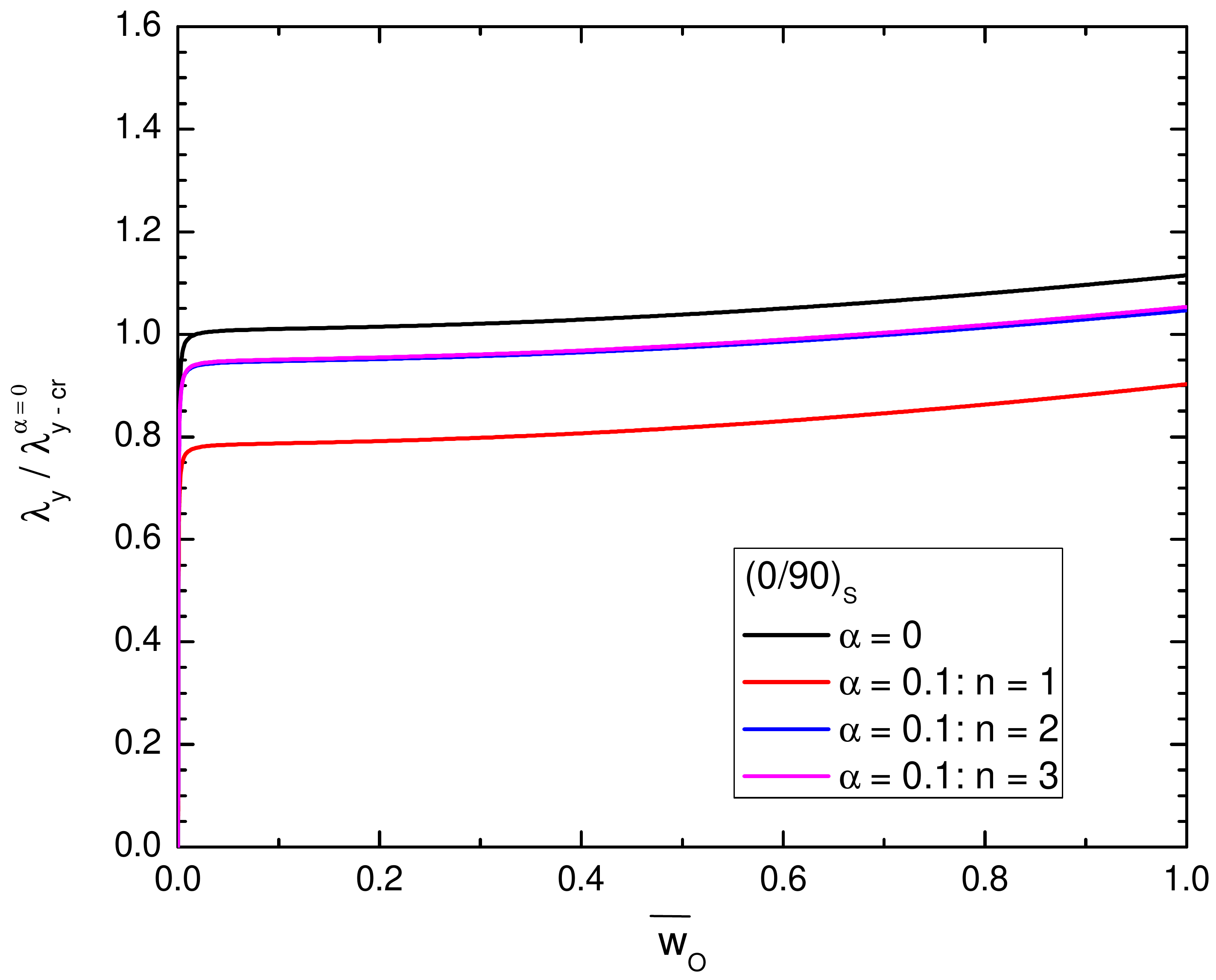}
\caption{
Buckling and postbuckling path of the isotropic square plate of sine-wave thickness with fixed amplitude $\lambda$ = 0.1 and varying the wavelength $n$ under uniaxial compression load in $y$ direction.}
\end{figure}

\begin{flushleft}
The effects of the amplitude parameter $\alpha$ and the number of wavelength $n$ to the critical load of the plates will be investigated. The obtained bifurcation behavior in nonlinear buckling analysis is given in \hyperref[Fig. 14]{Fig. 14} and \hyperref[Fig. 15]{Fig. 15} for the isotropic plate, and \hyperref[Fig. 16]{Fig. 16} and \hyperref[Fig. 17]{Fig. 17} for the cross-ply laminate. It can be remarked that $\alpha$ and $n$ slightly affect the buckling and postbuckling path of the isotropic plates while the buckling load of the cross-ply laminate significantly fluctuates depended on these parameters. In details, when the wavelength is fixed at $n=1$ and increasing the amplitude of sine wave, \hyperref[Fig. 16]{Fig. 16},  the critical load of the cross-ply laminate dramatically drops in comparison with the laminate of uniform thickness. On the other hand, when the amplitude of sine wave is fixed at $\alpha=0.1$ and changing the shape the plates by increasing the number of wavelength, buckling load of the cross-ply laminate decreases considerably ($n$=1), but it rises again and approaches a certain value close to the critical load in case of uniform thickness ($n$=2,3), see \hyperref[Fig. 17]{Fig. 17}.
\end{flushleft}

\section{Conclusions}

Nonlinear bending and buckling analysis of symmetric composite plates of variable thickness based on the framework of isogeometric analysis were presented in this work. Control thickness parameters were introduced to approximate the thickness which varies over the plates. This allows us model smooth non-uniform surfaces while using plate theories for analysis. Numerical examples were provided to illustrate the accuracy and applicability of the method as well as present some interesting features in nonlinear analysis of isotropic and laminated composite plates of variable thickness. According to the present method and obtained numerical solutions, some conclusions could be drawn:

\begin{flushleft}
(i) The approach successfully obtains nonlinear bending and buckling behavior of variable thickness composite plates. Numerical problems of isotropic and laminated composite plates with different kinds of thickness changes were defined, giving reliable solutions in comparison with other reference methods.
\end{flushleft}

\begin{flushleft}
(ii) Nonlinear analysis should be carried out in designing plates of variable thickness, especially composite laminates whose critical load could considerably decrease in comparison with ones of uniform thickness having the same volume.
\end{flushleft}

\newpage




\section*{References}
\bibliographystyle{model3-num-names}
\bibliography{biblio}

\begin{thebibliography}{48}
\providecommand{\natexlab}[1]{#1}
\providecommand{\url}[1]{\texttt{#1}}
\providecommand{\urlprefix}{URL }
\expandafter\ifx\csname urlstyle\endcsname\relax
  \providecommand{\doi}[1]{doi:\discretionary{}{}{}#1}\else
  \providecommand{\doi}{doi:\discretionary{}{}{}\begingroup
  \urlstyle{rm}\Url}\fi
\providecommand{\eprint}[2][]{\url{#2}}
\providecommand{\BIBand}{and}
\providecommand{\bibinfo}[2]{#2}
\ifx\xfnm\undefined \def\xfnm[#1]{\unskip,\space#1}\fi
\bibitem[{Fertis(1993)}]{Fertis(1993)}
\bibinfo{author}{Fertis\xfnm[ D.G.]}.
\newblock \bibinfo{title}{Nonlinear Mechanics}.
\newblock \bibinfo{publisher}{CRC Press}; \bibinfo{year}{1993}.
\bibitem[{Ohga et~al.(1995)Ohga, Shigematsu and Kawaguchi}]{Ohga(1995)}
\bibinfo{author}{Ohga\xfnm[ M.]}, \bibinfo{author}{Shigematsu\xfnm[ T.]},
  \bibinfo{author}{Kawaguchi\xfnm[ K.]}.
\newblock \bibinfo{title}{Buckling analysis of thin-walled members with
  variable thickness}.
\newblock \bibinfo{journal}{J Struct Eng}
  \bibinfo{year}{1995};\bibinfo{volume}{121(6)}:\bibinfo{pages}{919--924}.
\bibitem[{Timoshenko and Woinowsky-Krieger(1959)}]{Timoshenko(1959)}
\bibinfo{author}{Timoshenko\xfnm[ S.]},
  \bibinfo{author}{Woinowsky-Krieger\xfnm[ S.]}.
\newblock \bibinfo{title}{Theory of plates and shells}.
\newblock \bibinfo{publisher}{McGraw-hill}; \bibinfo{year}{1959}.
\bibitem[{Wu and Liu(2001)}]{Wu(2001)}
\bibinfo{author}{Wu\xfnm[ T.Y.]}, \bibinfo{author}{Liu\xfnm[ G.R.]}.
\newblock \bibinfo{title}{Free vibration analysis of circular plates with
  variable thickness by the generalized differential quadrature rule}.
\newblock \bibinfo{journal}{Thin-Walled Structures}
  \bibinfo{year}{2001};\bibinfo{volume}{38}:\bibinfo{pages}{7967--7980}.
\bibitem[{Eisenberger and Alexandrov(2003)}]{Eisenberger(2003)}
\bibinfo{author}{Eisenberger\xfnm[ M.]}, \bibinfo{author}{Alexandrov\xfnm[
  A.]}.
\newblock \bibinfo{title}{Buckling loads of variable thickness thin isotropic
  plates}.
\newblock \bibinfo{journal}{Thin-Walled Structures}
  \bibinfo{year}{2003};\bibinfo{volume}{41(9)}:\bibinfo{pages}{871--889}.
\bibitem[{Shufrin and Eisenberger(2005)}]{Shufri(2005)}
\bibinfo{author}{Shufrin\xfnm[ I.]}, \bibinfo{author}{Eisenberger\xfnm[ M.]}.
\newblock \bibinfo{title}{Stability of variable thickness shear deformable
  plates—first order and high order analyses}.
\newblock \bibinfo{journal}{Thin-Walled Structures}
  \bibinfo{year}{2005};\bibinfo{volume}{43(2)}:\bibinfo{pages}{189--207}.
\bibitem[{Ganesan and Liu(2008)}]{Ganesan(2008)}
\bibinfo{author}{Ganesan\xfnm[ R.]}, \bibinfo{author}{Liu\xfnm[ D.Y.]}.
\newblock \bibinfo{title}{Progressive failure and post-buckling response of
  tapered composite plates under uni-axial compression}.
\newblock \bibinfo{journal}{Compos Struct}
  \bibinfo{year}{2008};\bibinfo{volume}{82(2)}:\bibinfo{pages}{159--176}.
\bibitem[{Dinardo and Lagace(1989)}]{Dinardo(1989)}
\bibinfo{author}{Dinardo\xfnm[ M.T.]}, \bibinfo{author}{Lagace\xfnm[ P.A.]}.
\newblock \bibinfo{title}{Buckling and postbuckling of laminated composite
  plates with ply dropoffs}.
\newblock \bibinfo{journal}{AIAA Journal}
  \bibinfo{year}{1989};\bibinfo{volume}{27}:\bibinfo{pages}{1392--1398}.
\bibitem[{Varughese and Mukherjee(1997)}]{Varughese(1997)}
\bibinfo{author}{Varughese\xfnm[ B.]}, \bibinfo{author}{Mukherjee\xfnm[ A.]}.
\newblock \bibinfo{title}{A ply drop-off element for analysis of tapered
  laminated composites}.
\newblock \bibinfo{journal}{Compos Struct}
  \bibinfo{year}{1997};\bibinfo{volume}{39}:\bibinfo{pages}{123--144}.
\bibitem[{Hughes et~al.(2005)Hughes, Cottrell and Bazilevs}]{Hughes(2005)}
\bibinfo{author}{Hughes\xfnm[ T.J.R.]}, \bibinfo{author}{Cottrell\xfnm[ J.A.]},
  \bibinfo{author}{Bazilevs\xfnm[ Y.]}.
\newblock \bibinfo{title}{Isogeometric analysis: Cad, finite elements, nurbs,
  exact geometry and mesh refinement}.
\newblock \bibinfo{journal}{Comput Methods Appl Mech Eng}
  \bibinfo{year}{2005};\bibinfo{volume}{194}:\bibinfo{pages}{4135--4195}.
\bibitem[{Cottrell et~al.(2009)Cottrell, Hughes and Bazilevs}]{Cottrell(2009)}
\bibinfo{author}{Cottrell\xfnm[ J.A.]}, \bibinfo{author}{Hughes\xfnm[ T.J.R.]},
  \bibinfo{author}{Bazilevs\xfnm[ Y.]}.
\newblock \bibinfo{title}{Isogeometric analysis: toward integration of CAD and
  FEA}.
\newblock \bibinfo{publisher}{John Wiley and Sons}; \bibinfo{year}{2009}.
\bibitem[{Piegl and Tiller(1997)}]{Piegl(1997)}
\bibinfo{author}{Piegl\xfnm[ L.]}, \bibinfo{author}{Tiller\xfnm[ W.]}.
\newblock \bibinfo{title}{The NURBS book (Monographys in Visual Communication),
  2nd edition}.
\newblock \bibinfo{publisher}{Springer-Verlag}; \bibinfo{year}{1997}.
\bibitem[{Cottrell et~al.(2007)Cottrell, Hughes and Reali}]{Cottrell(2007)}
\bibinfo{author}{Cottrell\xfnm[ J.A.]}, \bibinfo{author}{Hughes\xfnm[ T.J.R.]},
  \bibinfo{author}{Reali\xfnm[ A.]}.
\newblock \bibinfo{title}{Studies of refinement and continuity in isogeometric
  structural analysis}.
\newblock \bibinfo{journal}{Comput Methods Appl Mech Eng}
  \bibinfo{year}{2007};\bibinfo{volume}{196}:\bibinfo{pages}{4160--4183}.
\bibitem[{Bazilevs et~al.(2010)Bazilevs, Calo, Cottrel, Evans, Hughes, Lipson
  et~al.}]{Bazilevs(2010)}
\bibinfo{author}{Bazilevs\xfnm[ Y.]}, \bibinfo{author}{Calo\xfnm[ V.M.]},
  \bibinfo{author}{Cottrel\xfnm[ J.A.]}, \bibinfo{author}{Evans\xfnm[ J.A.]},
  \bibinfo{author}{Hughes\xfnm[ T.J.R.]}, \bibinfo{author}{Lipson\xfnm[ S.]},
  et~al.
\newblock \bibinfo{title}{Isogeometric analysis using t-splines}.
\newblock \bibinfo{journal}{Comput Methods Appl Mech Eng}
  \bibinfo{year}{2010};\bibinfo{volume}{199}:\bibinfo{pages}{229--263}.
\bibitem[{Nguyen-Thanh et~al.(2011{\natexlab{a}})Nguyen-Thanh, Nguyen-Xuan,
  Bordas and Rabczuk}]{Nguyen-Thanha(2011)}
\bibinfo{author}{Nguyen-Thanh\xfnm[ N.]}, \bibinfo{author}{Nguyen-Xuan\xfnm[
  H.]}, \bibinfo{author}{Bordas\xfnm[ S.]}, \bibinfo{author}{Rabczuk\xfnm[
  T.]}.
\newblock \bibinfo{title}{Isogeometric analysis using polynomial splines over
  hierarchical t-meshes for two-dimensional elastic solids}.
\newblock \bibinfo{journal}{Comput Methods Appl Mech Eng}
  \bibinfo{year}{2011}{\natexlab{a}};\bibinfo{volume}{200}:\bibinfo{pages}{1892--1908}.
\bibitem[{Nguyen-Thanh et~al.(2011{\natexlab{b}})Nguyen-Thanh, Kiendl,
  Nguyen-Xuan, Wucher, Bletzinger, Bazilevs et~al.}]{Nguyen-Thanhb(2011)}
\bibinfo{author}{Nguyen-Thanh\xfnm[ N.]}, \bibinfo{author}{Kiendl\xfnm[ J.]},
  \bibinfo{author}{Nguyen-Xuan\xfnm[ H.]}, \bibinfo{author}{Wucher\xfnm[ R.]},
  \bibinfo{author}{Bletzinger\xfnm[ K.U.]}, \bibinfo{author}{Bazilevs\xfnm[
  Y.]}, et~al.
\newblock \bibinfo{title}{Rotation free isogeometric thin shell analysis using
  pht-splines}.
\newblock \bibinfo{journal}{Comput Methods Appl Mech Eng}
  \bibinfo{year}{2011}{\natexlab{b}};\bibinfo{volume}{200}:\bibinfo{pages}{3410--3424}.
\bibitem[{Kleiss et~al.(2012)Kleiss, Juttler and Zulehner}]{Kleiss(2012)}
\bibinfo{author}{Kleiss\xfnm[ S.K.]}, \bibinfo{author}{Juttler\xfnm[ B.]},
  \bibinfo{author}{Zulehner\xfnm[ W.]}.
\newblock \bibinfo{title}{Enhancing isogeometric analysis by a finite
  element-based local refinement strategy}.
\newblock \bibinfo{journal}{Comput Methods Appl Mech Eng}
  \bibinfo{year}{2012};\bibinfo{volume}{213-216}:\bibinfo{pages}{168--182}.
\bibitem[{Schillinger et~al.(2012)Schillinger, Dede, Scott, Evans, Borden, Rank
  et~al.}]{Schillinger(2012)}
\bibinfo{author}{Schillinger\xfnm[ D.]}, \bibinfo{author}{Dede\xfnm[ L.]},
  \bibinfo{author}{Scott\xfnm[ M.A.]}, \bibinfo{author}{Evans\xfnm[ J.A.]},
  \bibinfo{author}{Borden\xfnm[ M.J.]}, \bibinfo{author}{Rank\xfnm[ E.]},
  et~al.
\newblock \bibinfo{title}{An isogeometric design-through-analysis methodology
  based on adaptive hierarchical refinement of nurbs, immersed boundary
  methods, and t-spline cad surfaces}.
\newblock \bibinfo{journal}{Comput Methods Appl Mech Eng}
  \bibinfo{year}{2012};\bibinfo{volume}{249-252}:\bibinfo{pages}{116--150}.
\bibitem[{Hughes et~al.(2010)Hughes, Reali and Sangalli}]{Hughes(2010)}
\bibinfo{author}{Hughes\xfnm[ T.J.R.]}, \bibinfo{author}{Reali\xfnm[ A.]},
  \bibinfo{author}{Sangalli\xfnm[ G.]}.
\newblock \bibinfo{title}{Efficient quadrature for nurbs-based isogeometric
  analysis}.
\newblock \bibinfo{journal}{Comput Methods Appl Mech Eng}
  \bibinfo{year}{2010};\bibinfo{volume}{199}:\bibinfo{pages}{301--313}.
\bibitem[{Auricchio et~al.(2012)Auricchio, Calabro, Hughes, Reali and
  Sangalli}]{Auricchio(2012)}
\bibinfo{author}{Auricchio\xfnm[ F.]}, \bibinfo{author}{Calabro\xfnm[ F.]},
  \bibinfo{author}{Hughes\xfnm[ T.J.R.]}, \bibinfo{author}{Reali\xfnm[ A.]},
  \bibinfo{author}{Sangalli\xfnm[ G.]}.
\newblock \bibinfo{title}{A simple algorithm for obtaining nearly optimal
  quadrature rules for nurbs-based isogeometric analysis}.
\newblock \bibinfo{journal}{Comput Methods Appl Mech Eng}
  \bibinfo{year}{2012};\bibinfo{volume}{249-252}:\bibinfo{pages}{15--27}.
\bibitem[{Luu et~al.(2015)Luu, N.~I.~Kim and Lee}]{Luu(2015)}
\bibinfo{author}{Luu\xfnm[ A.T.]}, \bibinfo{author}{N.~I.~Kim\xfnm[ N.I.]},
  \bibinfo{author}{Lee\xfnm[ J.]}.
\newblock \bibinfo{title}{Nurbs-based isogeometric vibration analysis of
  generally laminated deep curved beams with variable curvature}.
\newblock \bibinfo{journal}{Compos Struct}
  \bibinfo{year}{2015};\bibinfo{volume}{119}:\bibinfo{pages}{150--165}.
\bibitem[{Kapoor and Kapania(2012)}]{Kapoor(2012)}
\bibinfo{author}{Kapoor\xfnm[ H.]}, \bibinfo{author}{Kapania\xfnm[ R.K.]}.
\newblock \bibinfo{title}{Geometrically nonlinear nurbs isogeometric finite
  element analysis of laminated composite plates}.
\newblock \bibinfo{journal}{Compos Struct}
  \bibinfo{year}{2012};\bibinfo{volume}{94}:\bibinfo{pages}{3434--3447}.
\bibitem[{Thai et~al.(2012)Thai, Nguyen-Xuan, Nguyen-Thanh, Le, Nguyen-Thoi and
  Rabczuk}]{Thai(2012)}
\bibinfo{author}{Thai\xfnm[ C.H.]}, \bibinfo{author}{Nguyen-Xuan\xfnm[ H.]},
  \bibinfo{author}{Nguyen-Thanh\xfnm[ N.]}, \bibinfo{author}{Le\xfnm[ T.H.]},
  \bibinfo{author}{Nguyen-Thoi\xfnm[ T.]}, \bibinfo{author}{Rabczuk\xfnm[ T.]}.
\newblock \bibinfo{title}{Static, free vibration, and buckling analysis of
  laminated composite reissner–mindlin plates using nurbs-based isogeometric
  approach}.
\newblock \bibinfo{journal}{Int J Numer Methods Eng}
  \bibinfo{year}{2012};\bibinfo{volume}{91}:\bibinfo{pages}{571--603}.
\bibitem[{Thai et~al.(2013)Thai, Ferreira, Carrera and
  Nguyen-Xuan}]{Thai(2013b)}
\bibinfo{author}{Thai\xfnm[ C.H.]}, \bibinfo{author}{Ferreira\xfnm[ A.J.M.]},
  \bibinfo{author}{Carrera\xfnm[ E.]}, \bibinfo{author}{Nguyen-Xuan\xfnm[ H.]}.
\newblock \bibinfo{title}{Isogeometric analysis of laminated composite and
  sandwich plates using a layerwise deformation theory}.
\newblock \bibinfo{journal}{Compos Struct}
  \bibinfo{year}{2013};\bibinfo{volume}{104}:\bibinfo{pages}{196--214}.
\bibitem[{Nguyen-Xuan et~al.(2013)Nguyen-Xuan, Thai and
  Nguyen-Thoi}]{Nguyen-Xuan(2013)}
\bibinfo{author}{Nguyen-Xuan\xfnm[ H.]}, \bibinfo{author}{Thai\xfnm[ C.H.]},
  \bibinfo{author}{Nguyen-Thoi\xfnm[ T.]}.
\newblock \bibinfo{title}{Isogeometric finite element analysis of composite
  sandwich plates using a higher order shear deformation theory}.
\newblock \bibinfo{journal}{Compos Part B}
  \bibinfo{year}{2013};\bibinfo{volume}{55}:\bibinfo{pages}{558--574}.
\bibitem[{Nguyen and Nguyen-Xuan(2013)}]{NguyenVP(2013)}
\bibinfo{author}{Nguyen\xfnm[ V.P.]}, \bibinfo{author}{Nguyen-Xuan\xfnm[ H.]}.
\newblock \bibinfo{title}{High-order b-splines based finite elements for
  delamination analysis of laminated composites}.
\newblock \bibinfo{journal}{Compos Struct}
  \bibinfo{year}{2013};\bibinfo{volume}{102}:\bibinfo{pages}{261--275}.
\bibitem[{Le-Manh and Lee(2014{\natexlab{a}})}]{Le-Manh(2014a)}
\bibinfo{author}{Le-Manh\xfnm[ T.]}, \bibinfo{author}{Lee\xfnm[ J.H.]}.
\newblock \bibinfo{title}{Postbuckling of laminated composite plates using
  nurbs-based isogeometric analysis}.
\newblock \bibinfo{journal}{Compos Struct}
  \bibinfo{year}{2014}{\natexlab{a}};\bibinfo{volume}{109}:\bibinfo{pages}{286--293}.
\bibitem[{Le-Manh and Lee(2014{\natexlab{b}})}]{Le-Manh(2014b)}
\bibinfo{author}{Le-Manh\xfnm[ T.]}, \bibinfo{author}{Lee\xfnm[ J.H.]}.
\newblock \bibinfo{title}{Stacking sequence optimization for maximum strengths
  of laminated composite plates using genetic algorithm and isogeometric
  analysis}.
\newblock \bibinfo{journal}{Compos Struct}
  \bibinfo{year}{2014}{\natexlab{b}};\bibinfo{volume}{116}:\bibinfo{pages}{357--363}.
\bibitem[{Guo et~al.(2014{\natexlab{a}})Guo, Nagy and Gürdal}]{Guo(2014a)}
\bibinfo{author}{Guo\xfnm[ Y.]}, \bibinfo{author}{Nagy\xfnm[ A.P.]},
  \bibinfo{author}{Gürdal\xfnm[ Z.]}.
\newblock \bibinfo{title}{A layerwise theory for laminated composites in the
  framework of isogeometric analysis}.
\newblock \bibinfo{journal}{Compos Struct}
  \bibinfo{year}{2014}{\natexlab{a}};\bibinfo{volume}{107}:\bibinfo{pages}{447--457}.
\bibitem[{Guo et~al.(2014{\natexlab{b}})Guo, Ruess and Gürdal}]{Guo(2014b)}
\bibinfo{author}{Guo\xfnm[ Y.]}, \bibinfo{author}{Ruess\xfnm[ M.]},
  \bibinfo{author}{Gürdal\xfnm[ Z.]}.
\newblock \bibinfo{title}{A contact extended isogeometric layerwise approach
  for the buckling analysis of delaminated composites}.
\newblock \bibinfo{journal}{Compos Struct}
  \bibinfo{year}{2014}{\natexlab{b}};\bibinfo{volume}{116}:\bibinfo{pages}{55--66}.
\bibitem[{Tran et~al.(2013{\natexlab{a}})Tran, Ferreira and
  Nguyen-Xuan}]{Tran(2013a)}
\bibinfo{author}{Tran\xfnm[ L.V.]}, \bibinfo{author}{Ferreira\xfnm[ A.J.M.]},
  \bibinfo{author}{Nguyen-Xuan\xfnm[ H.]}.
\newblock \bibinfo{title}{Isogeometric analysis of functionally graded plates
  using higher-order shear deformation theory}.
\newblock \bibinfo{journal}{Composites Part B: Engineering}
  \bibinfo{year}{2013}{\natexlab{a}};\bibinfo{volume}{51}:\bibinfo{pages}{368--383}.
\bibitem[{Tran et~al.(2013{\natexlab{b}})Tran, Thai and
  Nguyen-Xuan}]{Tran(2013b)}
\bibinfo{author}{Tran\xfnm[ L.V.]}, \bibinfo{author}{Thai\xfnm[ C.H.]},
  \bibinfo{author}{Nguyen-Xuan\xfnm[ H.]}.
\newblock \bibinfo{title}{An isogeometric finite element formulation for
  thermal buckling analysis of functionally graded plates}.
\newblock \bibinfo{journal}{Finite Elements in Analysis and Design}
  \bibinfo{year}{2013}{\natexlab{b}};\bibinfo{volume}{73}:\bibinfo{pages}{65--76}.
\bibitem[{Nguyen-Xuan et~al.(2014)Nguyen-Xuan, Tran, Thai, Kulasegaram and
  Bordas}]{Nguyen-Xuan(2014)}
\bibinfo{author}{Nguyen-Xuan\xfnm[ H.]}, \bibinfo{author}{Tran\xfnm[ L.V.]},
  \bibinfo{author}{Thai\xfnm[ C.H.]}, \bibinfo{author}{Kulasegaram\xfnm[ S.]},
  \bibinfo{author}{Bordas\xfnm[ S.P.A.]}.
\newblock \bibinfo{title}{Isogeometric analysis of functionally graded plates
  using a refined plate theory}.
\newblock \bibinfo{journal}{Composites Part B: Engineering}
  \bibinfo{year}{2014};\bibinfo{volume}{64}:\bibinfo{pages}{222--234}.
\bibitem[{Hosseini et~al.(2014)Hosseini, Remmers, Verhoosel and
  Borst}]{Hosseini(2014)}
\bibinfo{author}{Hosseini\xfnm[ S.]}, \bibinfo{author}{Remmers\xfnm[ J.J.]},
  \bibinfo{author}{Verhoosel\xfnm[ C.V.]}, \bibinfo{author}{Borst\xfnm[ R.D.]}.
\newblock \bibinfo{title}{An isogeometric continuum shell element for
  non-linear analysis}.
\newblock \bibinfo{journal}{Comput Methods Appl Mech Eng}
  \bibinfo{year}{2014};\bibinfo{volume}{271}:\bibinfo{pages}{1--22}.
\bibitem[{Kapoor et~al.(2013)Kapoor, Kapania and Soni}]{Kapoor(2013)}
\bibinfo{author}{Kapoor\xfnm[ H.]}, \bibinfo{author}{Kapania\xfnm[ R.K.]},
  \bibinfo{author}{Soni\xfnm[ S.R.]}.
\newblock \bibinfo{title}{Interlaminar stress calculation in composite and
  sandwich plates in nurbs isogeometric finite element analysis}.
\newblock \bibinfo{journal}{Compos Struct}
  \bibinfo{year}{2013};\bibinfo{volume}{106}:\bibinfo{pages}{537--548}.
\bibitem[{Nguyen-Thanh et~al.(2015)Nguyen-Thanh, Valizadeh, Nguyen,
  Nguyen-Xuan, Zhuang, Areias et~al.}]{Nguyen-Thanh(2015)}
\bibinfo{author}{Nguyen-Thanh\xfnm[ N.]}, \bibinfo{author}{Valizadeh\xfnm[
  N.]}, \bibinfo{author}{Nguyen\xfnm[ M.N.]},
  \bibinfo{author}{Nguyen-Xuan\xfnm[ H.]}, \bibinfo{author}{Zhuang\xfnm[ X.]},
  \bibinfo{author}{Areias\xfnm[ P.]}, et~al.
\newblock \bibinfo{title}{An extended isogeometric thin shell analysis based on
  kirchhoff–love theory}.
\newblock \bibinfo{journal}{Comput Methods Appl Mech Eng}
  \bibinfo{year}{2015};\bibinfo{volume}{284}:\bibinfo{pages}{265--291}.
\bibitem[{Kiendl et~al.(2009)Kiendl, Bletzinger, Linhard and
  Wuchner}]{Kiendl(2009)}
\bibinfo{author}{Kiendl\xfnm[ J.]}, \bibinfo{author}{Bletzinger\xfnm[ K.U.]},
  \bibinfo{author}{Linhard\xfnm[ J.]}, \bibinfo{author}{Wuchner\xfnm[ R.]}.
\newblock \bibinfo{title}{Isogeometric shell analysis with kirchhoff-love
  elements}.
\newblock \bibinfo{journal}{Comput Methods Appl Mech Eng}
  \bibinfo{year}{2009};\bibinfo{volume}{198}:\bibinfo{pages}{3902--3914}.
\bibitem[{Benson et~al.(2010)Benson, Bazilevs, Hsu and Hughes}]{Benson(2010)}
\bibinfo{author}{Benson\xfnm[ D.J.]}, \bibinfo{author}{Bazilevs\xfnm[ Y.]},
  \bibinfo{author}{Hsu\xfnm[ M.C.]}, \bibinfo{author}{Hughes\xfnm[ T.J.R.]}.
\newblock \bibinfo{title}{Isogeometric shell analysis: The reissner-mindlin
  shell}.
\newblock \bibinfo{journal}{Comput Methods Appl Mech Eng}
  \bibinfo{year}{2010};\bibinfo{volume}{199}:\bibinfo{pages}{276--289}.
\bibitem[{Benson et~al.(2011)Benson, Bazilevs, Hsu and Hughes}]{Benson(2011)}
\bibinfo{author}{Benson\xfnm[ D.J.]}, \bibinfo{author}{Bazilevs\xfnm[ Y.]},
  \bibinfo{author}{Hsu\xfnm[ M.C.]}, \bibinfo{author}{Hughes\xfnm[ T.J.R.]}.
\newblock \bibinfo{title}{A large deformation, rotation-free, isogeometric
  shell}.
\newblock \bibinfo{journal}{Comput Methods Appl Mech Eng}
  \bibinfo{year}{2011};\bibinfo{volume}{200}:\bibinfo{pages}{1367--13789}.
\bibitem[{Echter et~al.(2013)Echter, Oesterle and Bischoff}]{Echter(2013)}
\bibinfo{author}{Echter\xfnm[ R.]}, \bibinfo{author}{Oesterle\xfnm[ B.]},
  \bibinfo{author}{Bischoff\xfnm[ M.]}.
\newblock \bibinfo{title}{A hierarchic family of isogeometric shell finite
  elements}.
\newblock \bibinfo{journal}{Comput Methods Appl Mech Eng}
  \bibinfo{year}{2013};\bibinfo{volume}{254}:\bibinfo{pages}{170--180}.
\bibitem[{Kiendl et~al.(2015)Kiendl, F.~Auricchio and Reali.}]{Kiendl(2015)}
\bibinfo{author}{Kiendl\xfnm[ J.]}, \bibinfo{author}{F.~Auricchio L. Beirao
  da~Veiga\xfnm[ C.L.]}, \bibinfo{author}{Reali.\xfnm[ A.]}.
\newblock \bibinfo{title}{Isogeometric collocation methods for the
  reissner–mindlin plate problem}.
\newblock \bibinfo{journal}{Comput Methods Appl Mech Eng}
  \bibinfo{year}{2015};\bibinfo{volume}{284}:\bibinfo{pages}{489--507}.
\bibitem[{Deng et~al.(2015)Deng, Korobenko, Yan and Bazilevs}]{Deng(2015)}
\bibinfo{author}{Deng\xfnm[ X.]}, \bibinfo{author}{Korobenko\xfnm[ A.]},
  \bibinfo{author}{Yan\xfnm[ J.]}, \bibinfo{author}{Bazilevs\xfnm[ Y.]}.
\newblock \bibinfo{title}{Isogeometric analysis of continuum damage in
  rotation-free composite shells}.
\newblock \bibinfo{journal}{Comput Methods Appl Mech Eng}
  \bibinfo{year}{2015};\bibinfo{volume}{284}:\bibinfo{pages}{349--372}.
\bibitem[{Nguyen et~al.(2016)Nguyen, Thai and Nguyen-Xuan}]{BLieu(2016)}
\bibinfo{author}{Nguyen\xfnm[ L.B.]}, \bibinfo{author}{Thai\xfnm[ C.H.]},
  \bibinfo{author}{Nguyen-Xuan\xfnm[ H.]}.
\newblock \bibinfo{title}{A generalized unconstrained theory and isogeometric
  finite element analysis based on bezier extraction for laminated composite
  plates}.
\newblock \bibinfo{journal}{Eng Comput}
  \bibinfo{year}{2016};\bibinfo{volume}{32}:\bibinfo{pages}{1--19}.
\bibitem[{Tran et~al.(2016)Tran, Phung-Van, Lee, Wahab and
  Nguyen-Xuan}]{VLTran(2016)}
\bibinfo{author}{Tran\xfnm[ L.V.]}, \bibinfo{author}{Phung-Van\xfnm[ P.]},
  \bibinfo{author}{Lee\xfnm[ J.]}, \bibinfo{author}{Wahab\xfnm[ M.]},
  \bibinfo{author}{Nguyen-Xuan\xfnm[ H.]}.
\newblock \bibinfo{title}{Isogeometric analysis for nonlinear thermomechanical
  stability of functionally graded plates}.
\newblock \bibinfo{journal}{Compos Struct}
  \bibinfo{year}{2016};\bibinfo{volume}{140}:\bibinfo{pages}{655--667}.
\bibitem[{Reddy(2004{\natexlab{a}})}]{ReddyMechLaminates(2004)}
\bibinfo{author}{Reddy\xfnm[ J.N.]}.
\newblock \bibinfo{title}{Mechanics of laminated composite plates and shells,
  2nd edition}.
\newblock \bibinfo{publisher}{CRC Press}; \bibinfo{year}{2004}{\natexlab{a}}.
\bibitem[{Huu-Tai and Choi(2013)}]{HuuTai(2013)}
\bibinfo{author}{Huu-Tai\xfnm[ T.]}, \bibinfo{author}{Choi\xfnm[ D.]}.
\newblock \bibinfo{title}{A simple first-order shear deformation theory for
  laminated composite plates}.
\newblock \bibinfo{journal}{Compos Struct}
  \bibinfo{year}{2013};\bibinfo{volume}{106}:\bibinfo{pages}{754--763}.
\bibitem[{Reddy(2004{\natexlab{b}})}]{ReddyIntroNonlinear(2004)}
\bibinfo{author}{Reddy\xfnm[ J.N.]}.
\newblock \bibinfo{title}{An introduction to nonlinear finite element
  analysis}.
\newblock \bibinfo{publisher}{Oxford University Press, New York};
  \bibinfo{year}{2004}{\natexlab{b}}.
\bibitem[{Sundaresan et~al.(1996)Sundaresan, Singh and Rao}]{Sundaresan(1996)}
\bibinfo{author}{Sundaresan\xfnm[ P.]}, \bibinfo{author}{Singh\xfnm[ G.]},
  \bibinfo{author}{Rao\xfnm[ G.V.]}.
\newblock \bibinfo{title}{Buckling and postbuckling analysis of moderately
  thick laminated rectangular plates}.
\newblock \bibinfo{journal}{Comput Struct}
  \bibinfo{year}{1996};\bibinfo{volume}{61}:\bibinfo{pages}{79--86}.

\end{thebibliography}







\newpage

\section*{List of figures}

\newpage

\newpage

\newpage

\newpage

\newpage

\newpage

\newpage

\newpage

\newpage

\end{document}